\documentclass[pdflatex,sn-basic]{sn-jnl}% Basic Springer Nature Reference Style/Chemistry Reference Style

\usepackage{natbib}

\AtBeginDocument{%
}
\usepackage{hyperref}

\usepackage{lineno}
\usepackage{etoolbox}

\usepackage{graphics}%
\usepackage{multirow}%
\usepackage{amsmath,amssymb,amsfonts}%
\usepackage{amsthm}%
\usepackage{mathrsfs}%
\usepackage[title]{appendix}%
\usepackage[svgnames]{xcolor}
\usepackage{color, colortbl}

\usepackage{courier}
\definecolor{code}{HTML}{f7f7f7}

\usepackage{listings, microtype}

\lstdefinestyle{Common}
{   
	language={R},
	otherkeywords={1000, 10000,1:14},
	morekeywords={TRUE,FALSE},
	deletekeywords={data,frame,length,as,character,anova, sum, diag,log ,which, rep, factor, cbind, time, names, matrix, cov,c,pos, eigen, scale, shape, manual, expression, legend, list, labels, paste, text , axis, title, panel, grid, lm, formula, terms, index, col, round, na, rm, prop},
	frame=l,
	tabsize=2,
	framerule=3pt,
	framexleftmargin=4pt,
	framexrightmargin = 0.4pt,
	showstringspaces=false,
	morestring=*[b]",
}

\lstdefinestyle{A}
{
	style=Common,
	backgroundcolor = \color{code},
	basicstyle=\fontsize{9}{9}\ttfamily,
	keywordstyle=\color{blue},
	stringstyle=\color{DarkGreen},
	commentstyle=\color{DarkGreen},	
}

\usepackage{xpatch}% load the xpatch package
\xpatchbibdriver{misc}% patch the online driver:
{\printfield{entrysubtype}}% replace this line
{\printfield{entrysubtype}%  with these lines
	\newunit\newblock
	\printfield{howpublished}}
{}
{}

\usepackage{textcomp}%
\usepackage{manyfoot}%
\usepackage{booktabs}%
\usepackage{algorithm}%
\usepackage{algorithmicx}%
\usepackage{algpseudocode}%
\usepackage{listings}%
\usepackage{apacdoc}

\usepackage{tabularray}
\usepackage{adjustbox}
\usepackage{collcell}

\usepackage{graphicx}
\usepackage{array}
\newcolumntype{L}[1]{>{\raggedright\let\newline\\\arraybackslash}p{#1}}

\usepackage{rotating}
\usepackage{mathtools}
\usepackage{siunitx}

\usepackage{array}
\usepackage{arydshln}
\usepackage{boldline}
\usepackage{colortbl}

\usepackage{tikz}
\usetikzlibrary{arrows.meta,
	calc, chains,
	decorations.pathreplacing,%
	calligraphy,% had to be after decorations.pathreplacing
	shapes,
	shapes.geometric, arrows, matrix, positioning, fit, calc}
\definecolor{blue1}{RGB}{84,141,212}
\definecolor{blue2}{RGB}{142,180,227}
\definecolor{yellow1}{RGB}{255,229,153}
\definecolor{orange1}{RGB}{255,153,0}
\definecolor{gray1}{RGB}{127,127,127}
\definecolor{gray2}{rgb}{0.88,1,1}
\usepackage{bm}
\usepackage{dsfont}
\usepackage{pdflscape}
\usepackage{ltxtable, tabularx}
\usepackage{geometry}

\newcounter{alphasect}
\def\alphainsection{0}

\let\oldsection=\section
\def\section{%
	\ifnum\alphainsection=1%
	\addtocounter{alphasect}{1}
	\fi%
	\oldsection}%

\renewcommand\thesection{%
	\ifnum\alphainsection=1% 
	\Alph{alphasect}
	\else%
	\arabic{section}
	\fi%
}%

\newcolumntype{R}[2]{%
	>{\adjustbox{angle=#1,lap=\width-(#2)}\bgroup}%
	l%
	<{\egroup}%
}
\newcommand*\rot{\multicolumn{1}{R{30}{1.5em}}}% no optional argument here, please!

\usepackage{bbm}
\usepackage{float}

\usepackage{multicol}
\usepackage{subcaption}
\usepackage{caption}
\usepackage{stackengine}
\theoremstyle{thmstyleone}%
\theoremstyle{thmstyletwo}%

\theoremstyle{thmstylethree}%

\begin{document}

	\title[Article Title]{Descriptive Discriminant Analysis of Multivariate Repeated Measures Data: A Use Case}

% Commented author information	
	\author*[1]{\fnm{Ricarda} \sur{Graf}} \email{ricarda.graf@math.uni-augsburg.de}
	
	\author[2]{\fnm{Marina} \sur{Zeldovich}}
	
	\author[1]{\fnm{Sarah} \sur{Friedrich}}

	\affil[1]{\orgdiv{Department of Mathematics}, \orgname{University of Augsburg}, \city{Augsburg},  \country{Germany}}
	
	\affil[2]{\orgdiv{Institute of Medical Psychology and Medical Sociology}, \orgname{University Medical Center G{\"o}ttingen},  \city{G{\"o}ttingen},  \country{Germany}}

	\abstract{
Psychological research often focuses on examining group differences in a set of numeric variables for which normality is doubtful. Longitudinal studies enable the investigation of developmental trends. For instance, a recent study (\citealt{Voormolen2020}, \url{https://doi.org/10.3390/jcm9051525}) examined the relation of complicated and uncomplicated mild traumatic brain injury (mTBI) with multidimensional outcomes measured at three- and six-months after mTBI. The data were analyzed using robust repeated measures multivariate analysis of variance (MANOVA), resulting in significant differences between groups and across time points, then followed up by univariate ANOVAs per variable as is typically done. However, this approach ignores the multivariate aspect of the original analyses. We propose descriptive discriminant analysis (DDA) as an alternative, which is a robust multivariate technique recommended for examining significant MANOVA results and has not yet been applied to multivariate repeated measures data. We provide a tutorial with annotated \texttt{R} code demonstrating its application to these empirical data.

	}

	\keywords{descriptive discriminant analysis, MANOVA, multivariate repeated measures data, nonnormality, robustness, traumatic brain injury, variable ordering}

	\maketitle

	\section*{}
	
\begin{figure}[htb]
	\includegraphics[width=1.08\linewidth]{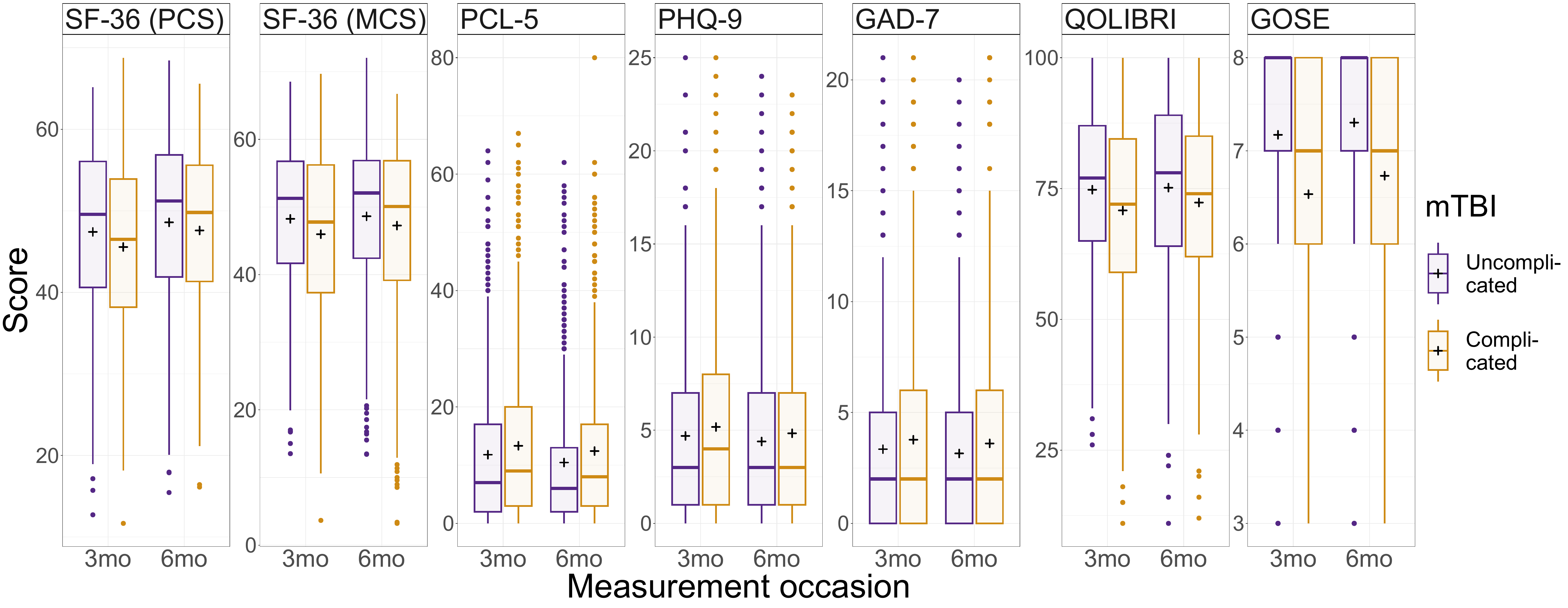}
	\caption{\footnotesize Boxplots showing the summary statistics of each variable per group (uncomplicated and complicated mild traumatic brain injury) and time point (3 and 6 months after TBI) in the CENTER-TBI data. Abbreviations: mTBI = mild traumatic brain injury;
		SF-PCS = Short Form (36) Health Survey (physical component score); SF-MCS = Short Form (36) Health Survey (mental component score);
		PCL-5 = Posttraumatic Stress Disorder Checklist; PHQ-9 = Patient Health Questionnaire; GAD-7 = Generalized Anxiety Disorder questionnaire; QOLIBRI = Quality of Life after Brain Injury; GOSE = Glasgow Outcome Scale-Extended; 3mo = 3 months; 6mo = 6 months.}
	\label{boxplot_orig_data}
\end{figure}

	 In their study, Voormolen et al. (\citeyear{Voormolen2020}) examined the association between two patient categories formed by patients having  experienced  either complicated or uncomplicated mild traumatic brain injury (mTBI), respectively, and a multidimensional outcome  based on data at three and six months after TBI. The multidimensional outcome comprised various clinical and psychological test scores. Data were obtained from the Collaborative European NeuroTrauma Effectiveness Research (CENTER-TBI) project 
		\citep{Maas2015, Maas2017}.\\ 
	 Voormolen et al. (\citeyear{Voormolen2020}) emphasized the need for research in the field due to high annual numbers of hospitalisations resulting from mild TBI. The longitudinal functional as well as cognitive differences in outcomes after uncomplicated and complicated mild TBI, respectively, were of particular interest  since researchers have come to contradictory conclusions in that regard. Computed tomography is used as a standard examination tool for diagnosing the complication of mTBI \citep{Williams1990}, which nowadays allows precise diagnoses. The outcome comprised seven clinical and psychological scores: The physical component summary (PCS) and the mental component summary (MCS) of the 36-item Short Form (SF-36v2) Health Survey \citep{Ware1992} and the 37-item Quality of Life after Brain Injury (QOLIBRI) instrument \citep{Steinbchel2010}, which are both self-report questionnaires assessing generic and disease-specific health-related quality of life, respectively. Furthermore, the Glasgow Outcome Scale-Extended (GOSE) for measuring functional recovery after TBI \citep{Jennett1981}, the Post-traumatic Stress Disorder Checklist-5 (PCL-5), a 20-item self-report questionnaire measuring symptoms of Post Traumatic Stress Disorder (PTSD) \citep{Blevins2015}, the Patient Health Questionnaire (PHQ-9), a nine-item self-report questionnaire evaluating depression symptoms experienced over the past two weeks \citep{Kroenke2002},  and the  Generalized Anxiety Disorder questionnaire (GAD-7), a seven-item self-report questionnaire assessing anxiety symptoms experienced over the past two weeks \citep{Spitzer2006}, were included as outcome measures. The sample included patients who completed all outcomes at both time points (three and six months after TBI), i.e. 569 patients with uncomplicated and 535 patients with complicated mild TBI, respectively. For further information on the study design and study sample, the interested reader is referred to Voormolen et al. (\citeyear{Voormolen2020}). Summary statistics of the dataset are visualized in Figure \ref{boxplot_orig_data}.\\  
	For statistical analysis,  Voormolen et al. (\citeyear{Voormolen2020}) chose a multivariate repeated-measures approach (MANOVA-RM) to screen variables for differences between patient groups, time points and possible interactions between these two effects \citep{Friedrich20181, Friedrich2022}, which was then followed up by multiple univariate analyses (ANOVA-RM), thus relinquishing the attempt of analyzing the combined influence of multiple correlated variables, overlooking the advantage of a multivariate follow-up analysis. \\
	Follow-up questions resulting from significant MANOVA findings are: How can the multivariate results be interpreted? Which variables contribute to the group differences and which time points matter the most?     
The usually recommended post-hoc technique to answer these questions is descriptive DA (e.g. \citealt{Huberty1989, Maxwell1992, Huberty2000, Bird2014}), a robust multivariate method for assessing the relative contribution of each variable to group separation. We propose using descriptive DA also in case of repeated measures and demonstrate its application to the empirical data set by Voormolen et al. (\citeyear{Voormolen2020}) providing the \texttt{R} code.

\section{Multivariate Analysis of Variance (MANOVA)}

\subsection{MANOVA in Psychological Research}
Multivariate methods are of particular importance in educational and psychological research where measurements of multiple  (continuous) correlated outcome variables are typically compared among  groups (categorical variables). There is usually no rationale for prioritizing single variables. Furthermore, variables are often measured by using scores corresponding to sums or averages of multiple measurements taken on a Likert scale. These scores are non-normally distributed due to the confined number of possible values \citep{Warner2013}, requiring methods suitable for non-normally distributed data.\\ 
MANOVA became available to applied researchers in the 1960s  \citep{Cooley1962, Cramer1966} and it is still one of the most commonly used statistical methods applied in the social sciences as well as other fields  \citep{Warne2012}. On the other hand, appropriate multivariate follow-up methods seem to be relatively unknown and are rarely applied. \\
Despite extensive methodological discussions, univariate follow-up techniques  for examining significant MANOVA effects prevail. Table \ref{reviews_MANOVA} gives an overview about some reviews of the social science literature that examined the frequency of uni- and multivariate  post-hoc techniques for MANOVA.  \\
Separate univariate F tests were among the first methods recommended for the analysis of group differences \citep{Cramer1966, Leary1980} and also described in textbooks \citep{Stevens1996, Tabachnick1996}, but the methodological literature clearly began opposing the MANOVA-ANOVAs approach \citep{Enders2003}. Keselman et al. (\citeyear{Keselman1998}) pointed out that in 84\% of the cases where MANOVA was applied,  conclusions were based on results obtained from separate univariate ANOVAs after Bonferroni correction for the type 1 error, an observation also made by Huang (\citeyear{Huang2020}).  The only reason to conduct the MANOVA was in fact a perceived additional control for type 1 error as promoted in the aforementioned sources. \\
Since MANOVA tests the null hypothesis of no difference among group means regarding the combined effect of  several correlated variables, the result cannot be compared to multiple independent results obtained from univariate tests  which ignore any dependencies between the variables. Due to the  inherent characteristics of behavioral science variables, multiple ANOVAs will likely lead to redundant results \citep{Huberty1989}. Multivariate analyses techniques will increase the sensitivity of detecting a particular variables' impact on group separation that may not be detected in univariate analyses \citep{Gnanadesikan1984} if the variables are correlated. Also, the idea of an additional type 1 error protection is a misconception since the type 1 error rates are  only maintained under circumstances that are not given in these applications \citep{Maxwell1992}. In particular, the type 1 error rates will only be maintained in case the  MANOVA null hypothesis either holds for all the variables (type 1 error will not exceed 5\%), is completely false (type 1 error cannot occur) or is false for all but one variable. Due to the different nature of multivariate and univariate analysis methods, a significant MANOVA effect does not necessarily imply any significant ANOVA effect \citep{Huberty1989}. \\
Since there is no theoretical support for the MANOVA-ANOVAs approach, other authors have also warned against it 
\citep{Thompson1999, Huberty2000, Enders2003, Bird2014}. Their advice is that researchers should rather decide at the beginning of their study whether uni- or multivariate effects are of interest and follow through with this initial plan.\\
Huberty and Smith (\citeyear{Huberty1982}) first supported the idea of multivariate follow-up analyses of significant MANOVA effects using descriptive discriminant analysis (DDA). Since then,  further sources besides the literature reviews listed in Table \ref{reviews_MANOVA}  have encouraged researchers in the social sciences to use descriptive DA as a more appropriate approach to examine group differences when considering a combination of correlated variables \citep{Sherry2006, Warne2014, Barton2016, Pituch2016}.\\
Here, we would like to demonstrate the application of descriptive DA in non-normally distributed multivariate repeated measures data using a real data example. 

\subsection{Repeated-Measures MANOVA}

In general, we would like to analyze measurements of $p$ variables  taken at $t$ time points for each individual $j = 1,..,n_i$ (where $\sum_{i=1}^g n_i = N$) in each group  $i = 1, \dots , g$. The vector $\mathbf{X}_{ij} = \{\mathbf{X}_{ij1}^T, \dots, \mathbf{X}_{ijt}^T \} \in \mathds{R}^{pt \times 1}$ contains the observations of the $j$th individual in the $i$th group, where $\mathbf{X}_{ijk} \in \mathds{R}^{p \times 1}$ is the vector of observations at time $k = 1,\dots,t$. The within group covariance matrices are denoted as $\bm{\Sigma}_i \in \mathds{R}^{pt \times pt}$ and the group means as $\bm{\mu}_{i} = (\bm{\mu}_{i1}^T, \dots, \bm{\mu}_{it}^T)^T \in \mathds{R}^{pt}$. The group variable   $i = 1, \dots , g$ represents the between-subject factor and the time variable $k = 1,\dots,t$ the within-subject factor.\\
Friedrich and Pauly (\citeyear{Friedrich20182}) propose a method for potentially nonnormally distributed data with unequal group covariance matrices for the general MANOVA model
\begin{equation}
	\mathbf{X}_{ij} = \bm{\mu}_i + \bm{\varepsilon}_{ij},
\end{equation}
which is suitable for a multivariate outcome at a single time point.\\
A factorial structure for multivariate repeated measures data can be incorporated in this model and is described in the appendix of Voormolen et al. (\citeyear{Voormolen2020}). It is implemented in the \texttt{R} package \texttt{MANOVA.RM} \citep{Friedrich2022}. This MANOVA model extended to multivariate repeated measures data is also suitable for nonnormally distributed data and assumes that group means $\bm{\mu}_i$ and within-group covariance matrices $\bm{\Sigma}_i$ exist. The  within-group covariance matrices may have any structure, and they may also be dissimilar (heteroscedastic). \\
The null hypotheses are formulated with respect to the group means $\bm{\mu} = (\bm{\mu}_1^T, \dots , \bm{\mu}_g^T)^T \in \mathds{R}^{gpt}$:
\begin{equation}
	H_0: \mathbf{T} \bm{\mu} = 0
\end{equation}
where $\mathbf{T}$ is a suitable contrast matrix. In particular, the main and interaction effects are tested by:
	\begin{align*}
		H_0 &: \mathbf{T}_G = \mathbf{P}_g \otimes \frac{1}{t} \mathbf{J}_t \otimes \mathbf{I}_p \qquad \text{(no group effect)} \\
		H_0 &: \mathbf{T}_T = \frac{1}{g} \mathbf{J}_g \otimes \mathbf{P}_t \otimes \mathbf{I}_p \qquad \text{(no time effect)} \\
		H_0 &: \mathbf{T}_{GT} = \mathbf{P}_g \otimes \mathbf{P}_t \otimes \mathbf{I}_p \qquad \text{(no interaction between group and time)}
	\end{align*}
where  $\mathbf{I}_p$ is the $p \times p$ identity matrix, $\mathbf{J}_t = \mathbf{1} \mathbf{1}^T$ is the $t \times t$ matrix of ones, and $\mathbf{P}_t = \mathbf{I}_t - \frac{1}{t} \mathbf{J}_t$ the $t \times t$ centering matrix, $\otimes$ the Kronecker product.\\
The modified ANOVA-type statistic (MATS) proposed by Friedrich and Pauly (\citeyear{Friedrich20182}) can still be applied in case of singular covariance matrices $\bm{\widehat{\Sigma}}_N = \text{diag}(N \bm{\widehat{\Sigma}}_i/ n_i)$ and it is scale-invariant. It is given by:
\begin{equation}
	Q_N = N \mathbf{\overline{X}}.^T \mathbf{T} (\mathbf{T}   \mathbf{\widehat{D}}_N     \mathbf{T})^+ \mathbf{T} \mathbf{\overline{X}}.
\end{equation}
where $\mathbf{\widehat{D}}_N = \text{diag}(N/n_i \cdot \hat{\sigma}^2_{iks}), \quad  i = 1, \dots , g, \, k = 1,\dots,t, \, s = 1,\dots,p$, $\mathbf{T}$ the hypothesis matrix, and  $\mathbf{\overline{X}}. = (\overline{X}_{1 \cdot 1}, \dots, \overline{X}_{g \cdot t})^T \in \mathds{R}^{gpt}$ the estimate of $\bm{\mu} \in \mathds{R}^{gpt}$.\\
$P$-values are calculated based on a bootstrap approach \citep{Friedrich20182}. For a large number $B$ of bootstrap samples, the test statistics $Q_N^{*,1}, \dots, Q_N^{*,B}$ are computed and the $p$-value is then defined as 
\begin{equation}
	p = \frac{1}{B} \sum\limits_{b = 1}^B \mathbbm{1}(Q_N \leq Q_N^{*,b}).
\end{equation}
Different bootstrap procedures \citep{Konietschke2015} that improve the inference for small sample sizes can be used, which are also implemented in the \texttt{R} package \texttt{MANOVA.RM} \citep{Friedrich2022}.

\begin{sidewaystable}[ph!]	
					
	\bgroup
	\def\arraystretch{3.5}
	
	\setlength\tabcolsep{2.5pt}

	\caption{\footnotesize Some reviews of the social science literature which examined the types of post-hoc techniques used to explain significant MANOVA results.}
	\begin{tabular}[t]{L{4.25cm}L{6.5cm}L{2cm}L{3.5cm}L{3.5cm}L{2.75cm} } 
		\rule{0pt}{2.5\normalbaselineskip} Reference   & Journal(s)/Database(s)	& Time span    &  \# (sign.) MANOVAs  & \# post-hoc: ANOVAs & \# post-hoc: DDA \\  \hline 
		\rule{0pt}{2.5\normalbaselineskip} \citeauthor{Huberty1989} \citeyear{Huberty1989} & 
		\parbox{21em}{\vspace{0.1cm}{\tiny$\bullet$} 6 behavioral science journals:\\
		Journal of Applied Psychology \\
		Journal of Counseling Psychology \\
		Journal of Consulting and Clinical Psychology \\
		Developmental Psychology \\
		Journal of Educational Psychology \\
		American Educational Research Journal}  & 1986 & 91 & 88 &  \parbox{10em}{ 2/88 additional DDA\\ 2/3 only DDA}   \\   \arrayrulecolor{lightgray}\hline 
		\rule{0pt}{2.5\normalbaselineskip} \citeauthor{Keselman1998} \citeyear{Keselman1998} & \parbox{21em}{{\tiny$\bullet$} 17 educational and behavioral science journals}  & 1994-1995 & 79& 66 & \parbox{8em}{ 4, but only 1 with substantive interpretation }\\ \arrayrulecolor{lightgray}\hline 
		\rule{0pt}{2.5\normalbaselineskip} \citeauthor{Kieffer2001} \citeyear{Kieffer2001} & \parbox{21em}{American Educational Research Journal (AERJ)\\
			Journal of Consumer Psychology (JCP)}  & 1988-1997 & \parbox{12em}{AERJ: 29 (multivariate)\\
		JCP: 160 (multivariate) } & \parbox{12em}{AERJ: 21/29 (univariate)\\
		JCP: 124/160 (univariate)} & infrequently \\  \arrayrulecolor{lightgray}\hline 
		\rule{0pt}{2.5\normalbaselineskip} \citeauthor{Armstrong2005} \citeyear{Armstrong2005} & \parbox{21em}{International Journal of Play Therapy}  & 1993-2003 & 3 & 2 & NA \\  \arrayrulecolor{lightgray}\hline 	
		\rule{0pt}{2.5\normalbaselineskip} \citeauthor{Warne2012} \citeyear{Warne2012} & \parbox{21em}{{\tiny$\bullet$} 5 gifted education research journals \\
		Gifted Child Quarterly  \\
		High Ability Studies \\
		Journal of Advanced Academics \\
		Journal for the Education of the Gifted  \\
		Roeper Review}  & 2006-2010 & 64 & 31 & 5\\  \arrayrulecolor{lightgray}\hline 
	
		\rule{0pt}{2.5\normalbaselineskip} \citeauthor{Tonidandel2013} \citeyear{Tonidandel2013} & \parbox{21em}{Journal of Applied Psychology}  & last three years & NA & 96\% & NA \\  \arrayrulecolor{lightgray}\hline 
	
		\rule{0pt}{2.5\normalbaselineskip} \citeauthor{Bird2014} \citeyear{Bird2014} & \parbox{21em}{APA PsycARTICLES, keyword "manova"\\
		included: random subset of 100 articles }  & 2001-2010 & 100 & 96 &  \parbox{10em}{3/96 additional DDA\\ 2/4 only DDA}\\  \arrayrulecolor{lightgray}\hline 
	
		\rule{0pt}{2.5\normalbaselineskip} \citeauthor{Warne2014} \citeyear{Warne2014} & \parbox{21em}{{\tiny$\bullet$} 3 psychology journals\\
		Journal of Clinical Psychology\\
		Emotion\\
		Journal of Counseling Psychology}  & 2009-2013 & 58 & NA & 0 \\   \arrayrulecolor{lightgray}\hline

		\rule{0pt}{2.5\normalbaselineskip} \citeauthor{AlAbdullatif2019} \citeyear{AlAbdullatif2019} & \parbox{21em}{{\tiny$\bullet$} 5 databases, 177 journals\\
		ERIC\\
		PsychINFO\\
		Academic Search Premier \\
		Professional Development Collection\\ 
		Teacher Reference Center\\
		included: any study reporting MANOVA together with any follow-up analysis
		}& 2013-2018  & 235 & 146 & 5  \\	\hline 
							      
	\end{tabular}

	\label{reviews_MANOVA}
	\egroup

\end{sidewaystable}

\section{Descriptive Discriminant Analysis (DDA)}\label{DDA}
\begin{figure}[htb]
	\centering

	\includegraphics[width=\linewidth]{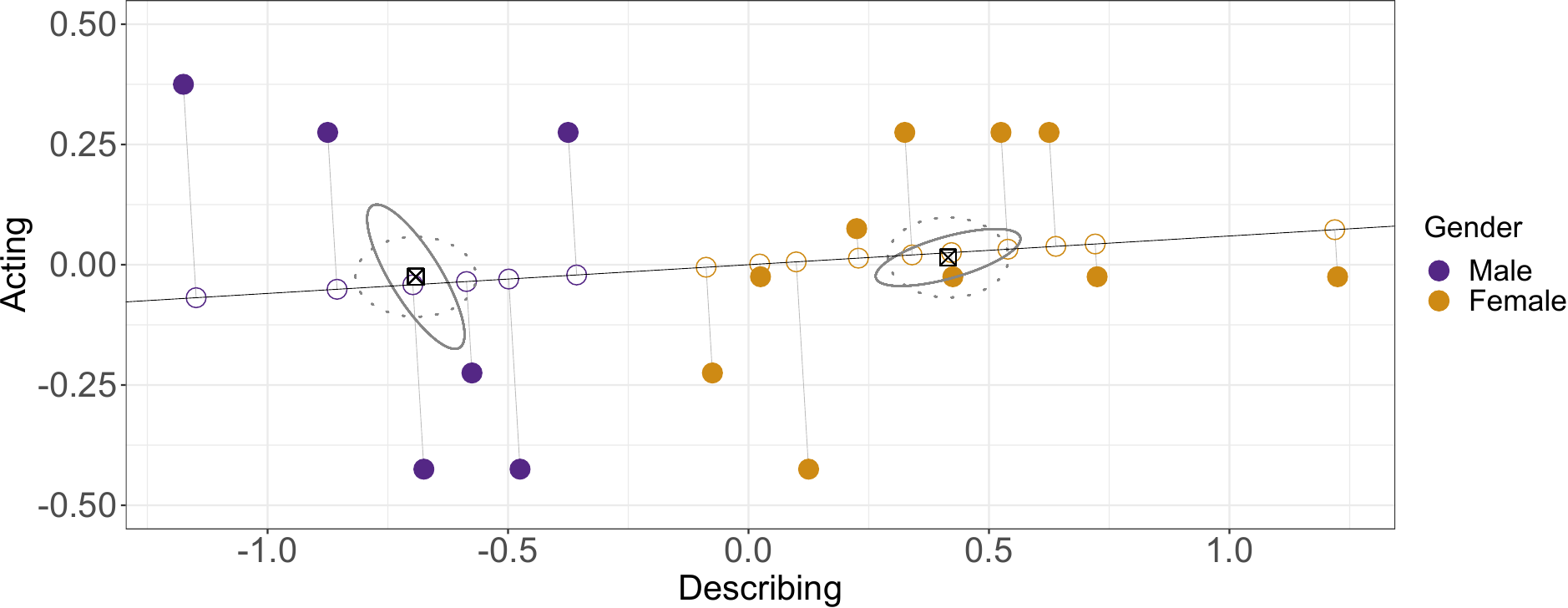}
	\caption{\footnotesize Illustrative 2D example of Fisher (descriptive) discriminant analysis: The actual data samples are shown as points in bold, the projected samples (Fisher discriminant function scores) are shown as circles, the actual (unequal) group covariances, $\bm{\widehat{\Sigma}}_{\text{Male}}$ and $\bm{\widehat{\Sigma}}_{\text{Female}}$, are shown as solid ellipses. From the pooled covariance matrix $\bm{\widehat{\Sigma}}_P$, indicated as a dashed ellipse, and the actual group means, indicated as crosses, the Fisher discriminant function is computed  (solid line).}   
	\label{Fisher_example}
\end{figure}

The term (linear) discriminant analysis comprises two approaches: predictive discriminant analysis (PDA) and descriptive  discriminant analysis (DDA). While DDA can be used for variable ordering as Fisher (\citeyear{Fisher1936}) demonstrated
in his original paper, PDA is used to specify observations as a member of one of multiple groups, which is an extension of Fisher's idea invented by Rao (\citeyear{Rao1948,Rao1973}) and based on normal theory. Descriptive DA does not make any assumption other than homogeneity of the covariance matrices, although statistical tests used in this context may require multivariate normality of the data.\\
Some authors point out that DDA requires a minimum sample size in order to obtain stable estimates of the pooled covariance matrix \citep{Huberty1975, Barcikowski1975}, i.e. interpretation should be done with caution unless the ratio of total sample size to number of variables is large, e.g. 20 to 1. Several authors found that multicollinearity may considerably affect the interpretation of descriptive DA coefficients \citep{Wilkinson1975, Borgen1978, Finn1978}. 
Two types of coefficients are available in descriptive DA, standardized discriminant function coefficients (DFCs) and structure coefficients. Uncertainty measures for these coefficients are not available, only the obtained point estimates can be interpreted \citep{Pituch2016}. \\ 
The within group covariance matrices $\bm{\Sigma}_i \in \mathds{R}^{pt \times pt}$   are assumed to be equal, i.e. $\bm{\Sigma}_i = \bm{\Sigma}_W \text{\quad for all \quad} i \in \{1, \dots , g\}$  (homoscedasticity), and non-singular (absence of multicollinearity). The between-group covariance matrix is denoted by $\bm{\Sigma}_B$.\\
In its original version, descriptive DA considers measurements taken at a single time point ($t = 1$) and determines a linear function of $p$ measurements ($X_{ij11} := X_{1} \in \mathds{R}, \dots, X_{ij1p} := X_{p} \in \mathds{R}$ for an arbitrary but fixed combination of $i \in \{1,\dots,g\}$ and $j \in \{1,\dots,n_i\}$, i.e. $N$ vectors ($X_{1}, \dots, X_{p}$) exist), also called Fisher discriminant function, which maximizes the ratio of between- to within-group variation \citep{Fisher1936, Venables2010}:
\begin{equation}\label{eq1}
	d = \lambda_1 X_{1} + \dots + \lambda_p X_{p}, \quad s.t. \quad \max\limits_{\bm{\lambda} \in \mathds{R}^p} \frac{\bm{\lambda}^T \bm{\Sigma}_B  \bm{\lambda}}{\bm{\lambda}^T  \bm{\Sigma}_W  \bm{\lambda}}
\end{equation}
where $\bm{\lambda} \in \mathds{R}^p$ represents the nonstandardized (or raw) DFCs and $d$ the Fisher discriminant scores, i.e. projections of the original measurements onto the linear discriminant function. The vector of DFCs best separating two groups ($i_1$ and $i_2$) can be estimated by 
\begin{equation}\label{eq2}
	\bm{\widehat{\lambda}} = \bm{\widehat{\Sigma}}_P^{-1} (\bm{\widehat{\mu}}_{i_1} - \bm{\widehat{\mu}}_{i_2}), \text{\quad where \quad} \bm{\widehat{\Sigma}}_P =  \frac{(n_{i_1} - 1) \bm{\widehat{\Sigma}}_{i_1}    + (n_{i_2} - 1) \bm{\widehat{\Sigma}}_{i_2}}{n_{i_1} + n_{i_2} - 2} 
\end{equation}
Standardized DFCs are  products of each variables nonstandardized (or raw) DFCs with their respective standard deviation, and represent the variable's association with the discriminant function holding the effects of all other variables constant. Variables associated with the largest absolute value coefficient contribute most to the group difference. The use of standardized DFCs has been recommended \citep{Pituch2016, Rencher1992, Finch2009, Finch2008} in contrast to structure coefficients. Structure coefficients are not recommended by these authors since they reflect only univariate information and frequently result in incorrect decisions.\\ 
In Figure \ref{Fisher_example} we have visualized a simple 2D example in order to explain the basic terms and  concept of descriptive discriminant analysis. We used a small subset of observations from "The Kentucky Inventory of Mindfulness Skills" dataset \citep{Baer2004, Baer2004data}.  We extracted  scores on the two scales "describing" and "acting" ($p=2$) of six male (purple) and ten female (yellow) participants for whom observations are made at a single time point ($t = 1$). The original observations are shown as points in bold. Through determining the optimal discriminant function (equation \ref{eq1} or \ref{eq2}, respectively), we have found the nonstandardized vector of DFCs,  $\bm{\lambda}$, and obtain the projections $d$, which are shown as circles. The Fisher discriminant function is the line passing through these projections. In this case, the group covariance matrices, $\bm{\widehat{\Sigma}}_{\text{Male}}$ and $\bm{\widehat{\Sigma}}_{\text{Female}}$ (shown as ellipses with solid lines), are not equal, rather  the correlation between the two variables has different signs. The pooled covariance matrix, $\bm{\widehat{\Sigma}}_P$ (shown as ellipse with dashed lines), which is the weighted average of the within-group covariances, and the group means (indicated by the crosses) are also shown.

\noindent Fisher discriminant analysis was developed for multivariate data measured at a single time point \citep{Fisher1936}, has also been applied to univariate repeated measures data \citep{Sajobi2012} and can be applied to factorial data, where each observation is assigned to exactly one of the categories of each factor \citep{Warner2013}. To the best of our knowledge, descriptive discriminant analysis has not yet been applied to multivariate repeated measures data, where measurements of the same variable at different time points are regarded as two different but correlated variables.
	
	\section{Results and Discussion: Application of MANOVA-RM and DDA to the CENTER-TBI Data}
	In this section, we replicate the repeated measures (M)ANOVA results for the CENTER-TBI data as presented in Voormolen et al. (\citeyear{Voormolen2020}) and discuss discriminant analysis as an alternative to multiple independent
	repeated measures ANOVAs.
	
	\subsection{(M)ANOVA-RM}
Significant differences in outcomes between uncomplicated and complicated mTBI and between time points were found in MANOVA-RM at a significance level of $\alpha$= .05. The interaction between main effects
was not significant. Replicated values of the test statistics and respective $p$-values are shown in Table \ref{RMMANOVA}.\\
In  Voormolen et al. (\citeyear{Voormolen2020}), multiple ANOVA-RM were chosen as post-hoc technique to examine the significant MANOVA-RM results. Significant test results ($\alpha_{\text{adj}}$=.007) are shown in bold in Table \ref{RMMANOVA}. $P$-values may differ from the results in Voormolen et al. (\citeyear{Voormolen2020}) because the computation depends on the specific seed that is chosen but the test results are identical. Table \ref{ANOVA_sign} gives an overview about the significance of the ANOVA-RM test results shown in Table \ref{RMMANOVA}.
Additionally, mean values and standard deviations per group and time point are shown in Table \ref{ANOVA_mean_sd}. Values corresponding to significant differences found in ANOVA-RM are shown in bold. For more details, please see the original analysis \citep{Voormolen2020}.

\begin{table}[htb]
		\caption{ \footnotesize Replicated repeated measures (M)ANOVA results for the CENTER-TBI data analyzed in Voormolen et al. (\citeyear{Voormolen2020}).
		MATS = modified ANOVA-type statistic, between-subject factor = TBI severity (uncomplicated and complicated mTBI), within-subject factor = time (time points 3 and 6 months after TBI),
	$p$ = $p$-value based on parametric bootstrapping, bold $p$-values are significant at $\alpha$ = .05 for MANOVA-RM and $\alpha_{adj.}$ = .007 for ANOVA-RM, respectively. Abbreviations: mTBI = mild traumatic brain injury;
	SF-PCS = Short Form (36) Health Survey (physical component score); SF-MCS = Short Form (36) Health Survey (mental component score);
	PCL-5 = Posttraumatic Stress Disorder Checklist; PHQ-9 = Patient Health Questionnaire; GAD-7 = Generalized Anxiety Disorder questionnaire; QOLIBRI = Quality of Life after Brain Injury; GOSE = Glasgow Outcome Scale-Extended.}
\label{RMMANOVA}
	{\footnotesize
\begin{tabular}{cllllll} 
	&&& \\
		\textbf{Analysis}  & \textbf{Dependent variable}		& \textbf{Independent variable}    &  \textbf{MATS} & \textbf{df1} & \textbf{df2} &  \textbf{$p$-value}\\ \cmidrule(lr){1-7}
		MANOVA		  &	All seven outcomes	&	mTBI       &  197.538  & \textminus & \textminus & \textbf{\textless{}.001} \\
		RM &						&	Time points      	&   34.708 & \textminus & \textminus	& \textbf{\textless{}.001}  \\ 
		&						&	mTBI:Time points 	&  2.932 & \textminus & \textminus & .152     \\ \cmidrule(lr){1-7}
		
		&	SF-36 PCS			&	mTBI       		 		&  5.897 & 1 & 1365.422 & .015 \\
		&						&	Time points      		&  61.133 & 1 & \textminus & \textbf{\textless{}.001}  \\ 
		&						&	mTBI:Time points 		&  4.361 & 1 & \textminus & .037    \\ \cmidrule(lr){2-7}		
			 
		&	SF-36 MCS				&	mTBI       		 		&  7.879 & 1 & 1399.985 &\textbf{.005} \\
		&						&	Time points      		&  10.502 & 1 & \textminus & \textbf{.001}  \\ 
		&						&	mTBI:Time points 		&  3.058 & 1 & \textminus & .08    \\ \cmidrule(lr){2-7}	
				 								 
		& PCL-5					&	mTBI       		 		&  5.481 & 1 & 1388.071 & .019 \\
		&						&	Time points      		&  16.902 & 1 & \textminus & \textbf{\textless{}.001}  \\
		&						&	mTBI:Time points 		&  0.653 & 1 & \textminus &.419    \\ \cmidrule(lr){2-7}	
				 							  
		ANOVA		  &	PHQ-9					&	mTBI    &  2.632 & 1 & 1386.136 & .105 \\
		RM&						&	Time points      		&  9.075 & 1 & \textminus & \textbf{.003}  \\
		&						&	mTBI:Time points 		&  0.032 & 1 & \textminus & .858    \\ \cmidrule(lr){2-7}	
				 								  								
		&	GAD-7					&	mTBI       		 	&  3.216 & 1 & 1425.187 & .073 \\
		&						&	Time points      		&  3.137 & 1 & \textminus &.077  \\
		&						&	mTBI:Time points 		&  0.026 & 1 & \textminus &.872    \\\cmidrule(lr){2-7}		
			  						
		&	QOLIBRI					&	mTBI       		 		&  12.25 & 1 & 1337.174 &\textbf{\textless{}.001} \\
		&						&	Time points      		&  8.588 & 1 & \textminus & \textbf{.003}  \\
		&						&	mTBI:Time points 		&  2.980 & 1 & \textminus & .084  \\ \cmidrule(lr){2-7}		
										 
		&	GOSE				&	mTBI       		 		&   80.944 & 1 & 1444.067 & \textbf{\textless{}.001} \\
		&						&	Time points      		&  26.150 & 1 & \textminus & \textbf{\textless{}.001}  \\
		&						&	mTBI:Time points 		&  1.057 & 1 & \textminus &.304  \\  \cmidrule(lr){1-7}	 %\bottomrule	           
	\end{tabular}
}
\end{table}

\begin{table}[htb]
	\caption{ \footnotesize Significance of the ANOVA-RM group effect (uncomplicated and complicated mTBI), time effect (3 and 6 months after mTBI) and interaction between both effects. $++$ = significant for $\alpha_{\text{adj}}$ =  .007, $-$ = not significant. Abbreviations: mTBI = mild traumatic brain injury; SF-PCS = Short Form (36) Health Survey (physical component score); SF-MCS = Short Form (36) Health Survey (mental component score); PCL-5 = Posttraumatic Stress Disorder Checklist; PHQ-9 = Patient Health Questionnaire; GAD-7 = Generalized Anxiety Disorder questionnaire; QOLIBRI = Quality of Life after Brain Injury; GOSE = Glasgow Outcome Scale-Extended.}
	\label{ANOVA_sign}
	{\small
		\begin{tabular}{lccc} 
			\toprule
	
			\textbf{Dependent variable}		&  \textbf{mTBI severity}    &  \textbf{Time points} & \textbf{Interaction}  \\ \cmidrule(lr){1-4}									   
			SF-36 PCS			&  \textminus   & $++$ & \textminus \\ 		 
			SF-36 MCS			&  $++$   & $++$ & \textminus \\ 			 								 
			PCL-5				&  \textminus   & $++$ & \textminus \\ 			 							  
			PHQ-9				&  \textminus   & $++$ & \textminus \\ 				 								  								
			GAD-7				&  \textminus    & \textminus & \textminus \\  						
			QOLIBRI				& $++$    & $++$ & \textminus \\							 
			GOSE				& $++$    & $++$ & \textminus \\      
			\arrayrulecolor{black}\bottomrule     
		\end{tabular}
	}
\end{table}

\begin{table}[htb]
	\caption{ \footnotesize Mean (standard deviation) compared between groups (uncomplicated and complicated mTBI) and time points (3 and 6 months after mTBI). Differences found significant in RM-ANOVA are shown in bold. Abbreviations: mTBI = mild traumatic brain injury;
		SF-PCS = Short Form (36) Health Survey (physical component score); SF-MCS = Short Form (36) Health Survey (mental component score);
		PCL-5 = Posttraumatic Stress Disorder Checklist; PHQ-9 = Patient Health Questionnaire; GAD-7 = Generalized Anxiety Disorder questionnaire; QOLIBRI = Quality of Life after Brain Injury; GOSE = Glasgow Outcome Scale-Extended.}
	\label{ANOVA_mean_sd}
	{\small
		\begin{tabular}{lllll} 
		\toprule
		 & \multicolumn{2}{c}{mTBI severity}&  \multicolumn{2}{c}{Time points}\\ \cmidrule(rl){2-3} \cmidrule(rl){4-5}
		 Dependent variable		&  \multicolumn{1}{c}{Uncompl.}    &  \multicolumn{1}{c}{Compl.} & \multicolumn{1}{c}{3 mo.}    &  \multicolumn{1}{c}{6 mo.}\\ \cmidrule(lr){1-5}									   
				SF-36 PCS			&   47.99 (10.48)  & 46.57 (10.16) & \textbf{46.51 (10.43)} & \textbf{48.1 (10.21)}\\ %\arrayrulecolor{gray!50}\cmidrule(lr){1-5}			 
				SF-36 MCS			&   \textbf{48.43 (11.01)}  	& \textbf{46.61 (12.07)} & \textbf{47.14 (11.56)} & \textbf{47.95 (11.56)}\\ 		 								 
			 	PCL-5				&   11.13 (12.71)   & 12.88 (13.69) & \textbf{12.54 (13.44)} & \textbf{11.41 (12.98)}\\ %\arrayrulecolor{gray!50}\cmidrule(lr){1-5}				 							  
				PHQ-9				&   4.54 (4.91)    	& 5.01 (5.14)   & \textbf{4.93 (5.04)} & \textbf{4.61 (5.01)}\\ 			 								  								
				GAD-7				&   3.25 (4.14)    	& 3.69 (4.64)   & 3.55 (4.45) & 3.37 (4.34)\\	%\arrayrulecolor{gray!50}\cmidrule(lr){1-5}	  						
				QOLIBRI				&   \textbf{74.97 (16.82)}   & \textbf{71.56 (17.33)} & \textbf{72.84 (17.05)} & \textbf{73.79 (17.21)}\\ %\arrayrulecolor{gray!50}\cmidrule(lr){1-5}									 
				GOSE				&   \textbf{7.24 (1.08)}    	& \textbf{6.63 (1.37)}   & \textbf{6.86 (1.3)} & \textbf{7.03 (1.23)}\\ %\arrayrulecolor{black}\cmidrule(lr){1-5}      
				\arrayrulecolor{black}\bottomrule     
		\end{tabular}
	}
\end{table}

	\subsection{DDA as Alternative Post-Hoc Analysis}
For descriptive DA, we consider the seven outcome variables ($p$ = 7), each measured at two time points ($t$ = 2) as 14 distinct correlated variables measured in two groups ($g$ = 2). Thus, the concept of descriptive DA can simply be applied to repeated measures data as well. In contrast to univariate follow-up strategies, it takes the correlation between variables and time points into account. With descriptive DA, an understanding of the relative importance of each of the correlated variables, i.e. its relative contribution to the overall significant multivariate effect found in MANOVA-RM, can be obtained.\\
We have $n_1$ = 569 and $n_2$ = 535 patients with uncomplicated and complicated mTBI, respectively, and consider (within-group and pooled) covariance matrices $\bm{\Sigma} \in \mathds{R}^{14 \times 14}$ and group means $\bm{\mu}_1, \bm{\mu}_2 \in \mathds{R}^{14}$.\\
\subsubsection{Assessment of Descriptive DA Assumptions}
\begin{figure}[htb!]
		 \hspace{3cm} 
		\includegraphics[width=0.64\linewidth]{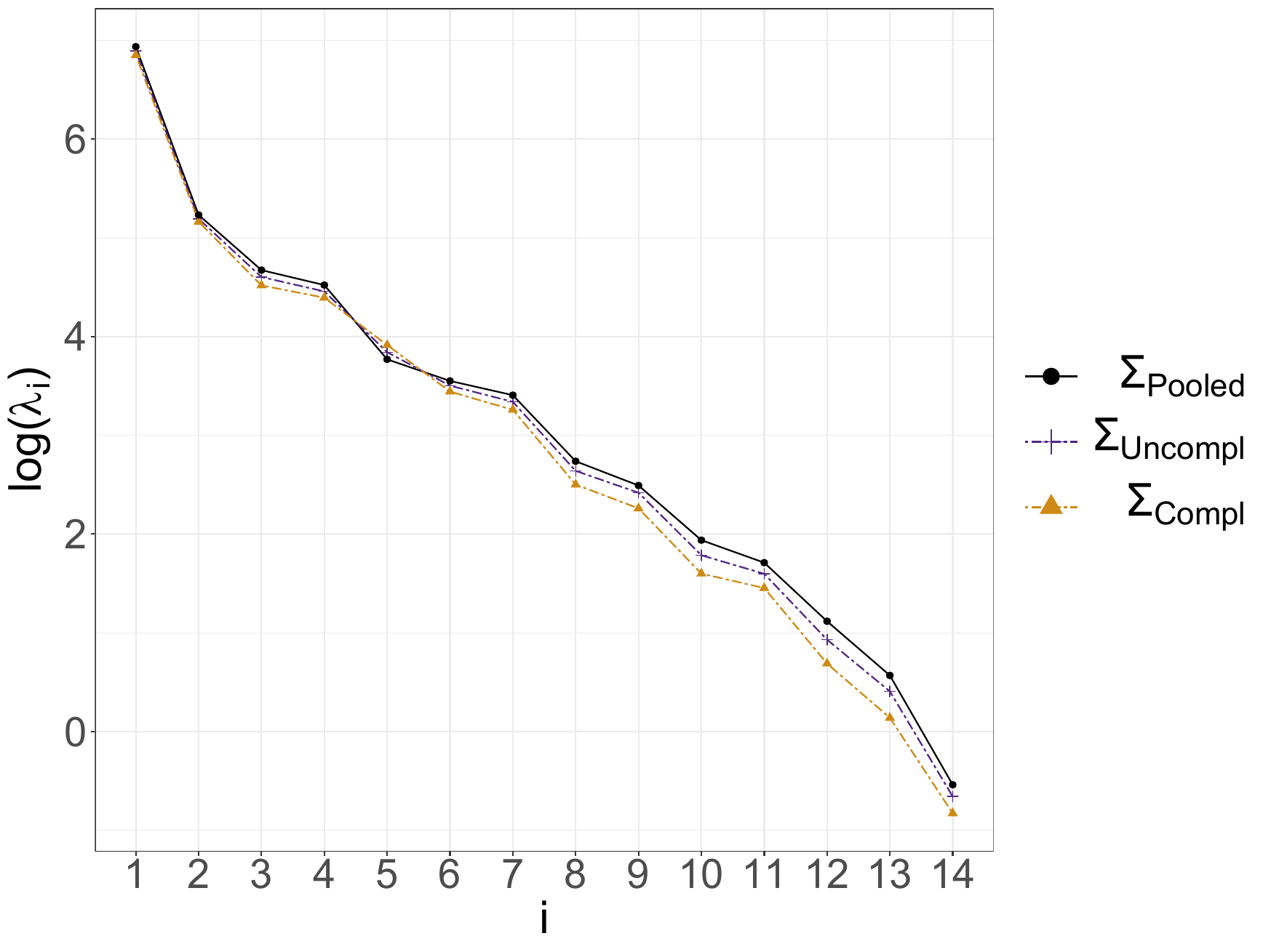}

		\caption{ \footnotesize {\raggedright Scree plot of log-eigenvalues of the within-group  covariance matrices  ($\bm{\Sigma}_{\text{Uncompl}}$ and $\bm{\Sigma}_{\text{Compl}}$) and the pooled covariance matrix ($\bm{\Sigma}_{\text{Pooled}}$) estimated from the CENTER-TBI dataset.}}
		\label{scree}
	\end{figure}

\begin{sidewaysfigure}[htb!]
	\includegraphics[width=\linewidth, height = 0.5\textheight]{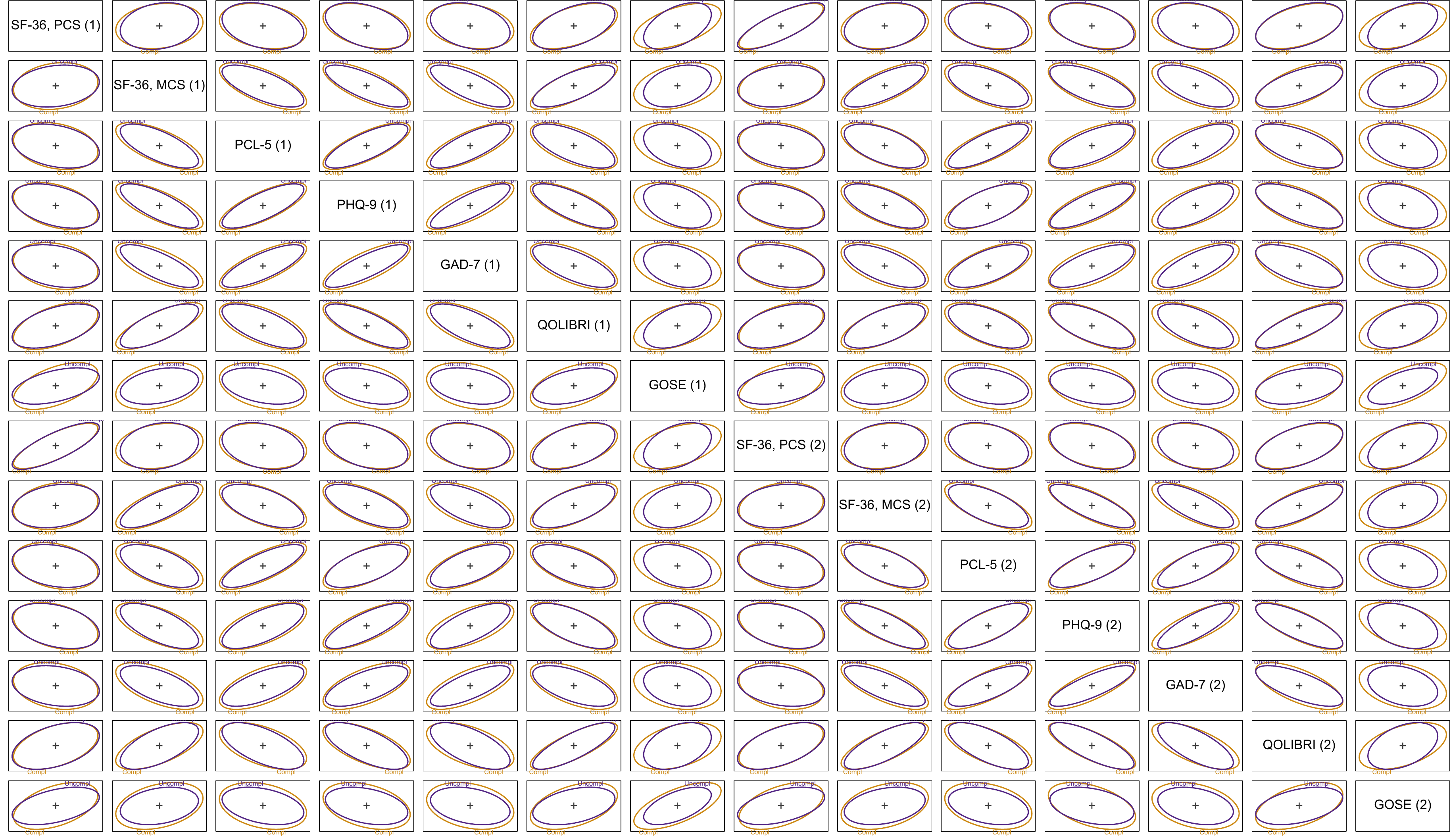}
	\caption{\footnotesize Plots of within-group covariance matrices for pairwise comparison of the variables in the CENTER-TBI dataset analyzed by Voormolen et al. (\citeyear{Voormolen2020}).}
	\label{pw_cov}
\end{sidewaysfigure}

\begin{figure}
	\raggedright
	\renewcommand\thefigure{S 1}
	\begin{subfigure}[b]{.32\linewidth}
		\includegraphics[width=\linewidth]{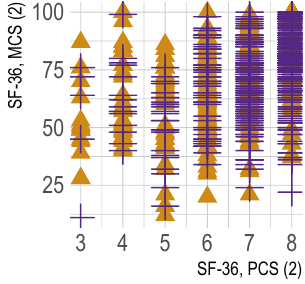}
	\end{subfigure}	
	\begin{subfigure}[]{.13\linewidth}
			\raggedright
		\includegraphics[width=\linewidth]{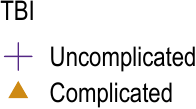}
		\vspace{3.5cm}
	\end{subfigure}	
		\caption{\small Scatterplots of each pair of variables present in the CENTER-TBI dataset analyzed by Voormolen et al. (\citeyear{Voormolen2020}) in order to examine equality of the within-group covariances. Observations belonging to patients with uncomplicated mTBI are indicated as purple crosses while observations of patients with complicated mTBI are shown as yellow triangles.}
		\label{Supp_scatter_cov}
\end{figure}

First, we will assess whether the assumption of homogeneity of (within-group) covariance matrices is fulfilled (i.e. if $\bm{\Sigma}_1 = \bm{\Sigma}_2$). Several strategies have been suggested.  The Box $M$ test \citep{Box1949} can be used to test the equality of $g$ covariance matrices, but it assumes multivariate normality and even a few outliers may strongly effect the test result. Another drawback is that the test is extremely powerful and inequality of the covariance matrices may not be identified even with a cutoff value   of $p<$.001 \citep{Huberty2000, Enders2003, Barton2016}. 
Since we assume that the data deviate from multivariate normality, we directly use alternative approaches. \\
Friendly and Sigal (\citeyear{Friendly2018}) suggest to compare scree plots showing the log-eigenvalues of the within-group covariance matrices and the pooled covariance matrix as an alternative to the Box M test. The Box M test considers the sum of differences between the log-determinants of each within-group covariance matrix and the pooled covariance matrix and evaluates their similarity. The log-determinant equals the sum of the log-eigenvalues.\\
Huberty and Lowman (\citeyear{Huberty2000}) proposed indices of generalized variance, i.e. to compare the equality of traces or equality of log-determinants, respectively, of the (within-class) covariances. If values are similar, the assumption of homogeneity can be assumed to be met.\\ 
Visual inspection using pairwise (2D) scatterplots of the data have also been suggested \citep{Barton2016} as well as pairwise comparison of the shape and direction of the within-group covariance matrices \citep{Friendly2018}. Using these visual tools, it may be difficult to come to a conclusion due to the potentially high number of pairwise comparisons. Figure \ref{Supp_scatter_cov} indicates similarity of the scatter plots for the two groups in the CENTER-TBI data set. \\
The scree plot (Figure \ref{scree}) shows approximate equality of the log-eigenvalues of each within-group covariance matrix compared to the log-eigenvalues of the pooled covariance matrix. Traces (sums of diagonal elements) of $\bm{\Sigma}_1$ and  $\bm{\Sigma}_2$ are 1434.8 and 1570.1, respectively. Log-determinants for both matrices are 39.4 and 42.1, respectively.  Equality of traces only evaluates the equality of variances among the groups, while log-determinants incorporate the covariances as well. Both measures have similar values for the CENTER-TBI dataset.
Figure \ref{pw_cov} shows the pairwise comparison of within-group covariances, where direction and shape of the ellipses appear to be overall comparable. Comparison of pairwise scatter plots of the data provide the same evidence and are shown in the supplementary Figure  \ref{Supp_scatter_cov}.  In total, we conclude that the assumption of homogeneous (within-group) covariance matrices is fulfilled.

\newpage
\noindent We will also examine if sample sizes are large enough \citep{Huberty1975, Barcikowski1975} and whether multicollinearity may complicate interpretation of the results \citep{Wilkinson1975, Borgen1978, Finn1978}, two factors that may influence the interpretability of the DFCs as mentioned above (Section \ref{DDA}). The ratio of total sample size to number of variables is 1104/14 = 78.9 $>$ 20 and thus  parameter estimates are assumed to be stable. There are several diagnostic tools for measuring multicollinearity, one of which is to inspect the scaled condition indices and their corresponding scaled variance decomposition proportions \citep{Belsley1980, Belsley1991} that can be arranged in a table for easy assessment. The standard approach, as described in Liao and Valliant (\citeyear{Liao2012}), is to check whether two or more large (scaled) variance decomposition proportions are associated with a (scaled) condition index, that is also large. Scaled condition indices of greater than 30 and scaled variance decomposition proportions of greater than .3 are usually considered as large. In combination, they may indicate the presence of multicollinearity between the respective variables. The results in Table \ref{multicoll1} indicate that there may be multicollinear variables in the CENTER-TBI dataset, since there are two (scaled) condition indices higher than 30 associated with at least two  (scaled) variance decomposition proportions higher than .3. Having determined the collinearity pattern, correlated variables are discarded one at a time and the collinearity measures are computed again. Unfortunately, in this case, variables can only be dropped simultaneously at both time points since complete data are required. \\
Different sets of variables can be dropped in order to eliminate multicollinearity. We did not consider excluding the GOSE score because it seems to be the most important variable, i.e. it has the highest association with group and time differences in the repeated measures ANOVA (Table \ref{RMMANOVA}) and the largest weights in descriptive DA in both cases, before and after removal of multicollinear variables (Table \ref{DFC1}, \ref{DFC_multcoll}, \ref{DFC_multcoll_suppl}). Only after removal of at least two of the three quality of life indices (QOLIBRI, SF-36 (MCS), SF-36 (PCS)) collinearity measures did not indicate multicollinearity anymore (Table \ref{multicoll2}, \ref{multicoll3}, \ref{multicoll4}).  
We will compare the descriptive DA results before and after removal of these potentially multicollinear variables. For comparison, we will repeat the same analyses for each time point (three and six months after mTBI) separately.

\begin{table}[htb!]
	\caption{ \footnotesize Assessment of multicollinearity using scaled condition indices and scaled variance decomposition proportions in the CENTER-TBI dataset. Only (scaled) variance decomposition proportions $>$ .3 are shown. The presence of (scaled) condition indices $>$ 30 indicate multicollinear variables if at least two of the associated (scaled) variance decomposition proportions are $>$ .3, without indicating which of the variables are of concern.}
	\label{multicoll1}
	{\footnotesize
		\begin{tabular}{L{1.1cm}ccccccccccccccccc} 
			\toprule
		\multirow{2}{*}[-1.2em]{\parbox{1.1cm}{\small{(Scaled)\\ condition\\ index}}}	& \multicolumn{14}{c}{\small{(Scaled) variance decomposition proportions}}& &&\\[1ex] \cmidrule(rl){2-18}
			& \rot{SF-36,PCS (1)} & \rot{SF-36,MCS (1)} & \rot{PCL-5 (1)} & \rot{PHQ-9 (1)} & \rot{GAD-7 (1)} & \rot{QOLIBRI (1)} & \rot{GOSE (1)} &

			\rot{SF-36,PCS (2)} & \rot{SF-36,MCS (2)} & \rot{PCL-5 (2)} & \rot{PHQ-9 (2)} & \rot{GAD-7 (2)} & \rot{QOLIBRI (2)} & \rot{GOSE (2)} & &&\\ \arrayrulecolor{gray!50}\cmidrule(lr){1-18} 
		2.18	 & . & . & . & . & . & . & . & . & . & . & . & . & . & . &  & &			 \\ \arrayrulecolor{gray!50}\cmidrule(lr){2-18}
		6.29	 & . & . & . & . & . & . & . & . & . & . & . & . & . & . & &  &			 \\ \arrayrulecolor{gray!50}\cmidrule(lr){2-18}	
		7.33	 & . & . & . & . & . & . & . & . & . & . & . & . & . & . & &  &			 \\ \arrayrulecolor{gray!50}\cmidrule(lr){2-18}
		8.21	 & . & . & . & . & . & . & . & . & . & . & . & . & . & . &  & &			 \\ \arrayrulecolor{gray!50}\cmidrule(lr){2-18}
		11.90	 & . & .  & .62 & . & . & . & . & . & . & .56 & . & . & . & . &&&			 \\ \arrayrulecolor{gray!50}\cmidrule(lr){2-18}
		13.34	  & . & . & . & .55 & . & . & . & . & . & . & .49 & .35 & . & . &&&			 \\ \arrayrulecolor{gray!50}\cmidrule(lr){2-18}
		14.11	 & . & . & . & . & . & . & . & . & . & . & . & . & . & . & &  &			 \\ \arrayrulecolor{gray!50}\cmidrule(lr){2-18} 
		20.64	 & . & . & . & . & . & . & . & . & . & . & . & . & . & . &  &  &			 \\ \arrayrulecolor{gray!50}\cmidrule(lr){2-18}
		28.19	 & . & . & . & . & . & . & . & . & . & . & . & . & . & . &  &  &			 \\ \arrayrulecolor{gray!95}\cmidrule(lr){1-18}	
		30.00	& . & . & . & . & . & . & . & . &  . & . & . & . & . & .37 &  &&			 \\ \arrayrulecolor{gray!50}\cmidrule(lr){2-18}
		31.04	& . & .48 & . & . & . & . & . & . & . & . & . & . & . & . &&&			 \\ \arrayrulecolor{gray!50}\cmidrule(lr){2-18}
		\textbf{36.28}	& . & . & . & . & . & . & \textbf{.47} & . & . & . & . & . & . & \textbf{.44} &&&			 \\ \arrayrulecolor{gray!50}\cmidrule(lr){2-18}
		47.54    & . & . & . & . & . & . & . & . & . & . & . & . & . & . &  &  &			 \\  \arrayrulecolor{gray!50}\cmidrule(lr){2-18} 
		\textbf{51.58}	& \textbf{.36} & . & . & . & . & \textbf{.4} & . & \textbf{.51} & \textbf{.48} & . & . & .  & \textbf{.45}  & . &&&			 \\ 				
			\arrayrulecolor{black}\bottomrule     
		\end{tabular}
	}
\end{table}

\begin{table}[htb!]
	\caption{ \footnotesize Assessment of multicollinearity using scaled condition indices and scaled variance decomposition proportions after removal of multicollinear variables (SF-36 PCS, QOLIBRI).  Only (scaled) variance decomposition proportions $>$ .3 are shown.}
	\label{multicoll2}
	{\footnotesize
		\begin{tabular}{L{2.3cm}cccccccccccc} 
			\toprule
			\multirow{2}{*}[-1.5em]{\parbox{2.3cm}{\small{(Scaled)\\ condition index}}}	& \multicolumn{10}{c}{\small{(Scaled) variance decomposition proportions}}& & \\[1ex] \cmidrule(rl){2-13}
			& \rot{SF-36, MCS (1)} & \rot{PCL-5 (1)} & \rot{PHQ-9 (1)} & \rot{GAD-7 (1)} & \rot{GOSE (1)}  & \rot{SF-36, MCS (2)} & \rot{PCL-5 (2)} & \rot{PHQ-9 (2)} & \rot{GAD-7 (2)} & \rot{GOSE (2)}    & &\\ \arrayrulecolor{gray!50}\cmidrule(lr){1-13} 
			2.1	 & . & . & . & . & . & . & . & . & . & . &  & 			 \\ \arrayrulecolor{gray!50}\cmidrule(lr){2-13}
			5.33	 & . & . & . & . & . & . & . & . & . & . &  &   			 \\ \arrayrulecolor{gray!50}\cmidrule(lr){2-13}	
			6.19	 & . & . & . & . & . & . & . & . & . & . &  &   			 \\ \arrayrulecolor{gray!50}\cmidrule(lr){2-13}
			6.99	 & . & . & . & . & . & . & . & . & . & .  &  &  			 \\ \arrayrulecolor{gray!50}\cmidrule(lr){2-13}
			10.09	 & . & .  & .66 & . & . & .61  & . & . & .& . &  &	 \\ \arrayrulecolor{gray!50}\cmidrule(lr){2-13}
			11.29	 & . & . & . & .69 & .32 & . & .32 &  . & .63 & .54 &  & 	 \\ \arrayrulecolor{gray!50}\cmidrule(lr){2-13}
			15.44	 & . & . & . & . & . & . & . & . & . & . &  &  		 \\ \arrayrulecolor{gray!50}\cmidrule(lr){2-13} 
			26.66	 & .35 & . & . & .44 & . & . & . & .42 & . & . &  &  		 \\ \arrayrulecolor{gray!50}\cmidrule(lr){2-13} 
			29.0	 & .48 & . & . & .35 & .44 & . & . & .37 & . & . &  &  		 \\ \arrayrulecolor{gray!95}\cmidrule(lr){1-13} 
			37.37	 & . & . & . & . & . & . & . & . & . & . &  &  		 \\				
			\arrayrulecolor{black}\bottomrule     
		\end{tabular}
	}
\end{table}

\begin{table}[htb!]
	\caption{ \footnotesize Assessment of multicollinearity using scaled condition indices and scaled variance decomposition proportions after removal of multicollinear variables (SF-36 PCS, SF-36 MCS).  Only (scaled) variance decomposition proportions $>$ .3 are shown.}
	\label{multicoll3}
	{\footnotesize
		\begin{tabular}{L{2.3cm}cccccccccccc} 
			\toprule
			\multirow{2}{*}[-1.5em]{\parbox{2.3cm}{\small{(Scaled)\\ condition index}}}	& \multicolumn{10}{c}{\small{(Scaled) variance decomposition proportions}}& & \\[1ex] \cmidrule(rl){2-13}
			 & \rot{PCL-5 (1)} & \rot{PHQ-9 (1)} & \rot{GAD-7 (1)} & \rot{QOLIBRI (1)}  &\rot{GOSE (1)}  & \rot{PCL-5 (2)} & \rot{PHQ-9 (2)} & \rot{GAD-7 (2)} & \rot{QOLIBRI (2)}  &\rot{GOSE (2)}    & &\\ \arrayrulecolor{gray!50}\cmidrule(lr){1-13} 
			2.11	 & . & . & . & . & . & . & . & . & . & . &  & 			 \\ \arrayrulecolor{gray!50}\cmidrule(lr){2-13}
			5.35	 & . & . & . & . & . & . & . & . & . & . &  &   			 \\ \arrayrulecolor{gray!50}\cmidrule(lr){2-13}	
			6.2	 & . & . & . & . & . & . & . & . & . & . &  &   			 \\ \arrayrulecolor{gray!50}\cmidrule(lr){2-13}
			6.98	 & . & . & . & . & . & . & . & . & . & .  &  &  			 \\ \arrayrulecolor{gray!50}\cmidrule(lr){2-13}
			10.1	 & .67 & .  & . & . & . & .59  & . & . & .& . &  &	 \\ \arrayrulecolor{gray!50}\cmidrule(lr){2-13}
			11.31	 & . & .63 & . & . & . & . & .32 &  . & .56 & . &  & 	 \\ \arrayrulecolor{gray!50}\cmidrule(lr){2-13}
			17.57	 & . & . & . & . & . & . & . & . & . & . &  &  		 \\ \arrayrulecolor{gray!50}\cmidrule(lr){2-13} 
			26.91	 & . & . & . & . & .61 & . & . & .6 & . & . &  &  		 \\ \arrayrulecolor{gray!95}\cmidrule(lr){1-13} 
			32.47	 & . & . & . & .66 & . & . & . & . & . & . &  &  		 \\ \arrayrulecolor{gray!50}\cmidrule(lr){2-13}
			33.76	 & . & . & . & . & . & . & . & . & . & .74 &  &  		 \\ 				
			\arrayrulecolor{black}\bottomrule     
		\end{tabular}
	}
\end{table}

\clearpage
\subsubsection{Application of Descriptive DA to Repeated Measures Data}

Table \ref{DFC1}, Table \ref{DFC_multcoll}, and Table \ref{DFC_multcoll_suppl} show the standardized discriptive discriminant coefficients (DFCs) before and after removal of potentially multicollinear variables from the CENTER-TBI dataset analyzed by Voormolen et al. (\citeyear{Voormolen2020}), respectively.\\
The DFCs for the entire set of variables (Table \ref{DFC1}) may be misleading because higher weights can be assigned randomly to one out of several highly correlated variables. Table \ref{DFC_multcoll} shows the DFCs after exclusion of redundant variables. Here, the high absolute values of the standardized DFCs coincide with the significant group and time effect of GOSE in ANOVA-RM (Table \ref{RMMANOVA}). QOLIBRI and SF-36 (MCS), the only other variables where both the time and group effect of the RM-ANOVA are significant, each have one DFC ranging after the highest DFCs corresponding to GOSE  (Table \ref{DFC_multcoll} (a) and (b)) indicating an influence of around 2/3 of that of GOSE. The other variables with either significant time or group effects, respectively, (SF-36 PCS, PCL-5, PHQ-9) still have higher DFCs compared to GAD-7, which does not have any significant RM-ANOVA effect.\\
The quality of life measures, SF-36 (PCS), SF-36 (MCS), and QOLIBRI, may reflect relations between the variables and severity of mTBI already included in other scores, and especially may contain repetitive information among each other since multicollinearity could only be removed after exclusion of at least two of the variables (according to the scaled condition indices and their respective scaled variance decomposition proportions).  Multicollinear variables should be removed before computing standardized DFCs because in case of multicollinearity,  relative weights  can be arbitrarily distributed among them. The exclusion of multicollinear variables does not affect the RM-MANOVA test results (Table \ref{RMMANOVA_coll}) - time and group effects remain significant.

\begin{table}[htb!]
		\caption{Standardized discriminant function coefficients (DFC) ordered by highest absolute value corresponding to the relative importance of the associated variable (original CENTER-TBI data)}
	\label{DFC1}
	\captionsetup{width=\textwidth}
	\begin{tabular}{ll}  \toprule
		Variable & Stand. DFC \\ \cmidrule(rl){1-2}
		GOSE (1)      & 0.6    \\ % x12
		GOSE (2)      & 0.58   \\ % x22
		SF-36, PCS (2)      & -0.47  \\ % x27
		SF-36, MCS (1)      & 0.38   \\ % x16
		PHQ-9 (1)      & 0.34   \\ % x14
		PCL-5 (2)      & -0.32  \\  % x23
		PCL-5 (1)      & 0.28   \\  % x13
		PHQ-9 (2)      & 0.23   \\  % x24
		SF-36, MCS (2)      & -0.23  \\  % x26
		QOLIBRI (1)      & 0.2    \\  % x15
		SF-36, PCS (1)      & 0.12   \\  % x17
		GAD-7 (2)      & -0.07  \\  % x21
		QOLIBRI (2)      & 0.05   \\  % QOLIBRI (2)
		GAD-7 (1)      & 0.02  % x11 
		\\ \arrayrulecolor{black}\bottomrule
	\end{tabular}
\end{table}

\begin{table}[htb!]
\caption{Standardized discriminant function coefficients (DFC) ordered by highest absolute value after removal of different sets of multicollinear variables.}
\label{DFC_multcoll}

		\begin{tabular}{ll} 
				
		\parbox{4cm}{(a) Removed variables:\\ SF-36 PCS, QOLIBRI}	& 
		\parbox{4cm}{(b) Removed variables:\\ SF-36 PCS, SF-36 MCS} \\
			\begin{tabular}{ll}  \toprule
					Variable & Stand. DFC \\ \cmidrule(rl){1-2}
					GOSE (1)      & -0.61    \\ % x17   
					GOSE (2)      & -0.53   \\ % x27    
					SF-36, MCS (1)      & -0.45  \\ % x12   
					PHQ-9 (2)       & -0.37   \\ % x24   
					PHQ-9 (1)      & -0.32   \\ % x14  
					PCL-5 (2)      & 0.28  \\  %  x23   
					PCL-5 (1)      & -0.2   \\  % x13   
					GAD-7 (2)     & 0.11   \\  % x25   
					SF-36, MCS (2)      & 0.09  \\  % x22   
					GAD-7 (1)      & -0.05  % x15
					\\ \arrayrulecolor{black}\bottomrule
			\end{tabular}
		& \hspace{-0.3cm}
			\begin{tabular}{ll}  \toprule
					Variable & Stand. DFC \\ \cmidrule(rl){1-2}
					GOSE (1)     & 0.6    \\ % x17     
					GOSE (2)     & 0.54   \\ % x27             
					QOLIBRI (1)      & 0.37  \\ % x16       
					PHQ-9 (2)     & 0.34   \\ %  x24          
					PCL-5 (2)     & -0.3   \\ % x23        
					PHQ-9 (1)     & 0.28  \\  %x14       
					QOLIBRI (2)      & -0.25   \\  % x26        
					PCL-5 (1)     & 0.17   \\  % x13       
					GAD-7 (2) & -0.15 \\ % x25     
					GAD-7 (1) & 0.0 \\ % x15			
					\arrayrulecolor{black}\bottomrule
			\end{tabular}
		\end{tabular}
\end{table}

\subsection{Descriptive DA per Time Point}

For comparison, we repeated the analyses for each time point separately in order to examine whether multicollinearity still occurs. Table \ref{multicoll_time1} and \ref{multicoll_time2} show that multicollinearity is not an issue for the analyses at a single time point but is rather introduced by repeated measurements and potentially higher order interactions in case of an increased number of variables.\\
Table \ref{DFC_per_time}  shows that GOSE is still the most important variable for distinguishing between patients with uncomplicated and complicated mild TBI, and GAD-7 is the least important variable at both time points. 

\begin{table}[htb!]
	\caption{ \footnotesize Assessment of multicollinearity using scaled condition indices and scaled variance decomposition proportions at 3 months after mTBI (time point 1).  Only (scaled) variance decomposition proportions $>$ .3 are shown.}
	\label{multicoll_time1}
	{\footnotesize
		\begin{tabular}{L{2.3cm}ccccccccc} 
			\toprule
			\multirow{2}{*}[-1.5em]{\parbox{2.3cm}{\small{(Scaled)\\ condition index}}}	& \multicolumn{9}{c}{\small{(Scaled) variance decomposition proportions}}\\[1ex] \cmidrule(rl){2-10}
			& \rot{SF-36, PCS (1)} &  \rot{SF-36, MCS (1)} & \rot{PCL-5 (1)}    & \rot{PHQ-9 (1)} & \rot{GAD-7 (1)} &  \rot{QOLIBRI (1)} & \rot{GOSE (1)} & &\\ \arrayrulecolor{gray!50}\cmidrule(lr){1-10} 
			2.18	 & . & . & . & . & . & . & . &   & 			 \\ \arrayrulecolor{gray!50}\cmidrule(lr){2-10}
			6.72	 & . & . & .85 & . & .42 & . & . &   &   			 \\ \arrayrulecolor{gray!50}\cmidrule(lr){2-10}	
			7.37	 & . & . & . & .67 & .45 & . & . &   &   			 \\ \arrayrulecolor{gray!50}\cmidrule(lr){2-10}
			12.95	 & .35 & . & . & . & . & . & . &   &  			 \\ \arrayrulecolor{gray!50}\cmidrule(lr){2-10}
			18.45	 & . & .  & . & . & . & .  & .89 &   &	 \\ \arrayrulecolor{gray!50}\cmidrule(lr){2-10}
			24.59	 & . & . & . & . & . & .88 & . &    & 	 \\ \arrayrulecolor{gray!50}\cmidrule(lr){2-10}
			31.57	 & . & .6 & . & . & . & . & . &  &  		 \\ 				
			\arrayrulecolor{black}\bottomrule     
		\end{tabular}
	}
\end{table}

\begin{table}[htb!]
	\caption{ \footnotesize Assessment of multicollinearity using scaled condition indices and scaled variance decomposition proportions at 6 months after mTBI (time point 2).  Only (scaled) variance decomposition proportions $>$ .3 are shown.}
	\label{multicoll_time2}
	{\footnotesize
		\begin{tabular}{L{2.3cm}ccccccccc} 
			\toprule
			\multirow{2}{*}[-1.5em]{\parbox{2.3cm}{\small{(Scaled)\\ condition index}}}	& \multicolumn{9}{c}{\small{(Scaled) variance decomposition proportions}} \\[1ex] \cmidrule(rl){2-10}
			& \rot{SF-36, PCS (2)} &  \rot{SF-36, MCS (2)} & \rot{PCL-5 (2)}    & \rot{PHQ-9 (2)} & \rot{GAD-7 (2)} &  \rot{QOLIBRI (2)} & \rot{GOSE (2)} & &\\ \arrayrulecolor{gray!50}\cmidrule(lr){1-10} 
			2.13	 & . & . & . & . & . & . & . &   & 			 \\ \arrayrulecolor{gray!50}\cmidrule(lr){2-10}
			6.31	 & . & . & .92 & . & . & . & . &   &   			 \\ \arrayrulecolor{gray!50}\cmidrule(lr){2-10}	
			7.19	 & . & . & . & .55 & .68 & . & . &   &   			 \\ \arrayrulecolor{gray!50}\cmidrule(lr){2-10}
			13.5	 & .35 & . & . & . & . & . & . &   &  			 \\ \arrayrulecolor{gray!50}\cmidrule(lr){2-10}
			18.76	 & . & .  & . & . & . & .  & .85 &   &	 \\ \arrayrulecolor{gray!50}\cmidrule(lr){2-10}
			26.15	 & .35 & . & . & . & . & .88 & . &    & 	 \\ \arrayrulecolor{gray!50}\cmidrule(lr){2-10}
			34.03	 & . & .57 & . & . & . & . & . &  &  		 \\ 				
			\arrayrulecolor{black}\bottomrule     
		\end{tabular}
	}
\end{table}

\begin{table}[htb!]
	\caption{ \footnotesize Standardized discriminant function coefficients (DFC) ordered by highest absolute value (separate analyses per time point).}
	\label{DFC_per_time}
	
	\begin{tabular}{lll} 
		
		\parbox{4cm}{(a) Time point 1\\ 
			(3 months after mTBI)}	& 
		\parbox{4cm}{(b) Time point 2\\ 
			(6 months after mTBI)} \\
		\begin{tabular}{ll}  \toprule
			Variable & Stand. DFC \\ \cmidrule(rl){1-2}
			GOSE (1)      & -1.01    \\ % x17   
			PHQ-9 (1)      & -0.53   \\ % x14        
			SF-36, MCS (1)  & -0.3  \\ % x12       
			QOLIBRI (1)       & -0.25   \\ % x16    
			SF-36, PCS (1)      & 0.14   \\ % x11     
			PCL-5 (1)      & -0.04  \\  % x13     
			GAD-7 (1)      & -0.03   \\  % x15    
			 \arrayrulecolor{black}\bottomrule
		\end{tabular}
		& \hspace{-0.3cm}
		\begin{tabular}{ll}  \toprule
			Variable & Stand. DFC \\ \cmidrule(rl){1-2}
			GOSE (2)      & -1.1    \\ % x27   
			PHQ-9 (2)      & -0.47   \\ % x24          
			SF-36, PCS (2)  & 0.35  \\ %  x21        
			PCL-5 (2)       & 0.27   \\ % x23        
			QOLIBRI (2)      & -0.23   \\ % x26        
			SF-36, MCS (2)      & 0.07  \\  % x22       
			GAD-7 (2)      & 0.01   \\  % x25 			
			\arrayrulecolor{black}\bottomrule
		\end{tabular}
	\end{tabular}
\end{table}

\clearpage
\subsection{R Code}
Assuming the data are arranged in long format (columns: group variable "Complicated", "timepoints", "id", and the seven measurements of psychological and clinical scores) in a data frame "data\underline{\hspace{0.175cm}}long", and in wide format
(columns: group variable "Complicated", 2 $\times$ 7 measurements of psychological and clinical scores) in a data frame "data\underline{\hspace{0.175cm}}wide", respectively, the following \texttt{R} code shows how the analyses from the previous sections can be performed in \texttt{R}.

% %  	`\newpage`
%\newpage
\begin{lstlisting}[style = A, flexiblecolumns=true, escapeinside=``,mathescape=true]
	
	library(MANOVA.RM)
	library(MASS)
	library(heplots)
	library(ggplot2)
	library(Rfast)
	library(candisc)
	library(mctest)
	
	# arrange data in long format and in wide format
	
	### Repeated measures MANOVA (Table 2)
	manova_tbi <-  multRM(cbind(SF36_PCS,SF36_MCS,PCL5,
															PHQ9,GAD7,
															QoLIBRI,GOSE) ~ Complicated * timepoints, 
															data = data_TBI_long, 
															subject = "id", 
															within = "timepoints", 
															iter = 10000, 
															resampling = "paramBS", 
															seed = 123)
															
	### Repeated measures ANOVA (Table 2)
	anova_tbi_SF36_PCS <-  RM(SF36_PCS ~ Complicated * timepoints, 
																		data = data_TBI_long, 
																		subject = "id", 
																		within = "timepoints", 
																		iter = 1000, 
																		resampling = "paramBS", 
																		seed = 123) 
	anova_tbi_SF36_MCS <-  RM(SF36_MCS ~ Complicated * timepoints, 
																		data = data_TBI_long, 
																		subject = "id", 
																		within = "timepoints", 
																		iter = 1000, 
																		resampling = "paramBS", 
																		seed = 123) 	
	anova_tbi_PCL5 <-  RM(PCL5 ~ Complicated * timepoints, 
															data = data_TBI_long, 
															subject = "id", 
															within = "timepoints", 
															iter = 1000, 
															resampling = "paramBS", 
															seed = 123)
	anova_tbi_PHQ9 <-  RM(PHQ9 ~ Complicated * timepoints, 
															data = data_TBI_long, 
															subject = "id", 
															within = "timepoints", 
															iter = 1000, 
															resampling = "paramBS", 
															seed = 123)
	anova_tbi_GAD7 <-  RM(GAD7 ~ Complicated * timepoints, 
															data = data_TBI_long, 
															subject = "id", 
															within = "timepoints", 
															iter = 1000, 
															resampling = "paramBS", 
															seed = 123)
	anova_tbi_QoLIBRI <-  RM(QoLIBRI ~ Complicated * timepoints, 
																	data = data_TBI_long, 
																	subject = "id", 
																	within = "timepoints", 
																	iter = 1000, 
																	resampling = "paramBS", 
																	seed = 123)
	anova_tbi_GOSE <-  RM(GOSE ~ Complicated * timepoints, 
															data = data_TBI_long, 	
															subject = "id", 
															within = "timepoints", 
															iter = 1000, 
															resampling = "paramBS", 
															seed = 123)				
														
	### Pairwise comparison of within-group covariances (Figure 4)
	heplots::covEllipses(x = data_wide[,c("SF-36 PCS (1)","SF-36 MCS (1)",
																							"PCL-5 (1)","PHQ-9 (1)","GAD-7 (1)",
																							"QOLIBRI (1)","GOSE (1)", "SF-36 PCS (2)",
																							"SF-36 MCS (2)","PCL-5 (2)","PHQ-9 (2)",
																							"GAD-7 (2)","QOLIBRI (2)","GOSE (2)")], 
											 				group = as.factor(data_wide$\text{\small{\$}}$Complicated), 
															fill = c(rep(FALSE,2), TRUE), variables=1:14, 
															fill.alpha = .1,
															center = TRUE,
															label.pos = c("top","bottom"),
															pooled = FALSE,
															col =  c("$\color{DarkGreen}{\#}$ce8a14", "$\color{DarkGreen}{\#}$542785"))
														
	### Indices of generalized variance (Section 4.2.1)
	
	# data per group
	data0 <- data_wide[which(data_wide$\text{\small{\$}}$Complicated == 0), 
												c("SF-36 PCS (1)","SF-36 MCS (1)",
												"PCL-5 (1)","PHQ-9 (1)","GAD-7 (1)",
												"QOLIBRI (1)","GOSE (1)", "SF-36 PCS (2)",
												"SF-36 MCS (2)","PCL-5 (2)","PHQ-9 (2)",
												"GAD-7 (2)","QOLIBRI (2)","GOSE (2)")]
	data1 <- data_wide[which(data_wide$\text{\small{\$}}$Complicated == 1), 
												c("SF-36 PCS (1)","SF-36 MCS (1)",
												"PCL-5 (1)","PHQ-9 (1)","GAD-7 (1)",
												"QOLIBRI (1)","GOSE (1)", "SF-36 PCS (2)",
												"SF-36 MCS (2)","PCL-5 (2)","PHQ-9 (2)",
												"GAD-7 (2)","QOLIBRI (2)","GOSE (2)")]
												
	# Traces of within-group cavariance matrices
	sum(diag(data0))
	sum(diag(data1))	
	
	# Log-determinants of within-group cavariance matrices
	log(det(data0))
	log(det(data1))     
	
	### Scree-Plot of log-eigenvalues (Figure 3)
	covp <- pooled.cov(as.matrix(cbind(data0, data1)), data_wide$\text{\small{\$}}$Complicated)
	cov0 <- cov(data0)
	cov1 <- cov(data1)
	
	log_eig <- data.frame(n = rep(c(1:14),3),
															g = c(rep("p",14), rep("u",14), rep("c",14)),
															e = c(log(eigen(covp)$\text{\small{\$}}$val),
																			log(eigen(cov0)$\text{\small{\$}}$val), 
																			log(eigen(cov1)$\text{\small{\$}}$val)))																		
	ggplot(log_eig, aes(x = n, y = e, group = g, color = g)) +
					geom_line(aes(linetype = g)) +
					xlab("i") +
					ylab(expression("log("*paste(lambda)[i]*")")) +
					guides(shape = guide_legend(override.aes = list(size = 5))) +
					theme_bw() +
					geom_point(data = log_eig, 
								aes(x = n, y = e, group = g, color = g, shape = g), 
								size = 2) +
					scale_color_manual(labels = c(expression("\u03A3"[Pooled]),
																									expression("\u03A3"[Uncompl]),
																									expression("\u03A3"[Compl])),
																		name = "",
																		values = c("black","$\color{DarkGreen}{\#}$542785", "$\color{DarkGreen}{\#}$ce8a14")) +
					scale_shape_manual(labels = c(expression("\u03A3"[Pooled]),
																									expression("\u03A3"[Uncompl]),
																									expression("\u03A3"[Compl])), 
																		name = "",
																		values = c(16,3, 17)) +
					scale_linetype_manual(labels = c(expression("\u03A3"[Pooled]),
																											expression("\u03A3"[Uncompl]),
																											expression("\u03A3"[Compl])),
																				name = "", 
																				values = c("solid","twodash", "twodash")) +
					scale_x_continuous(breaks = c(1:14)) +
					theme(legend.text = element_text(size = 30),
					legend.key.size = unit(3,"line"),
					axis.text = element_text(size = 27),
					axis.title = element_text(size = 30),
					panel.grid.minor.x = element_blank()) 
							
	### Multicollinearity assessment (Table 5)
	# scaled condition indices
	cond.index(formula = Complicated ~ SF-36 PCS (1) + SF-36 MCS (1) + 
															PCL-5 (1) + PHQ-9 (1) + GAD-7 (1) + 
															QOLIBRI (1) + GOSE (1) +  SF-36 PCS (2) + 
															SF-36 MCS (2) + PCL-5 (2) + PHQ-9 (2) + 
															GAD-7 (2) + QOLIBRI (2) + GOSE (2),  
								data = data_wide) %>% round(.,2)
	# scaled variance decomposition proportion							
	eigprop(mod = lm(Complicated ~ SF-36 PCS (1) + SF-36 MCS (1) + 
												PCL-5 (1) + PHQ-9 (1) + GAD-7 (1) + 
												QOLIBRI (1) + GOSE (1) +  SF-36 PCS (2) + 
												SF-36 MCS (2) + PCL-5 (2) + PHQ-9 (2) + 
												GAD-7 (2) + QOLIBRI (2) + GOSE (2), 
								data = data_wide), na.rm = FALSE, Inter = TRUE, prop = 0.5)$\text{\small{\$}}$pi
								
	### Discriminant function coefficients (DFC)
	# non-standardized DFC
	MASS::lda(Complicated~., data = data_wide[,c("Complicated", 
																	"SF-36 PCS (1)","SF-36 MCS (1)","PCL-5 (1)",
																	"PHQ-9 (1)","GAD-7 (1)","QOLIBRI (1)",
																	"GOSE (1)",	 "SF-36 PCS (2)","SF-36 MCS (2)",
																	"PCL-5 (2)","PHQ-9 (2)","GAD-7 (2)",
																	"QOLIBRI (2)","GOSE (2)")])$\text{\small{\$}}$scaling
	
	# standardized DFC (Table 9)
	x = lm(formula = cbind(SF-36 PCS (1) + SF-36 MCS (1) + 
													PCL-5 (1) + PHQ-9 (1) + GAD-7 (1) + 
													QOLIBRI (1) + GOSE (1) +  SF-36 PCS (2) + 
													SF-36 MCS (2) + PCL-5 (2) + PHQ-9 (2) + 
													GAD-7 (2) + QOLIBRI (2) + GOSE (2)) ~ Complicated, 
						data = data_wide)
	y = candisc(x, terms = "Groups")
	y$\text{\small{\$}}$coeffs.std
	
\end{lstlisting}
                
% Rao F test , Sajobi 2020
% https://search.r-project.org/CRAN/refmans/yacca/html/F.test.cca.html   (F.test.cca )
% https://people.stat.sc.edu/hitchcock/chapter8_R_examples.txt    (cancor2)

%\begin{figure}[htb]
%	\includegraphics[width=0.45\linewidth]{figures/Figure_corr_plot_tbi.pdf} 
%\end{figure}

\clearpage
	%% --------------------
	\section{Discussion}
	% --------------------	
	
	Longitudinal study designs are common in the social sciences. MANOVA-RM \citep{Friedrich20182, Voormolen2020}, a robust extension of MANOVA to repeated measures data, can be used even for non-normal data with unequal group-covariance matrices. However, meaningful post-hoc analyses following significant multivariate results are seldom used in the literature. \\
	In this tutorial, we demonstrated the application of descriptive DA to multivariate repeated measures data. This method has been suggested in methodological literature reviews as a more suitable post-hoc analysis to study significant MANOVA effects. \\
	Another possibility for judging (clinical) relevance of significant test results are effect sizes. For potentially non-normal multivariate repeated measures data, such effect measures are not yet available. A nonparametric effect size $A_w$ has been proposed for multivariate data measured at a single time point \citep{Li2021} but robust effect sizes measures for repeated measures data   are part of future research. \\ 
	Although descriptive DA only assumes equality of  group covariance matrices, estimates of the pooled covariance matrix required for computing the relative variable weights  may by unstable in case of certain deviations from multivariate normality: Sajobi et al. (\citeyear{Sajobi2012}) examined this effect in a simulation study for smaller sample sizes, i.e. some of their settings do not comply with the recommended ratio of total sample size to number of variables for descriptive DA \citep{Huberty1975, Barcikowski1975}. Sajobi et al. (\citeyear{Sajobi2012}) apply descriptive DA to repeated measures of a single variable and suggest more stable estimators of the covariance matrix. Possibly, this approach can be extended to estimate DFCs in case of deviations from multivariate normality in multivariate repeated measures data.\\
	More extensive simulation studies for special situations specific to multivariate repeated measures data might help in developing guidelines for the use of descriptive DA methods. \\
	Predictive discriminant analysis is often used in psychology to assess the discriminative ability of a set of variables in order to determine the usefulness of the particular set or in order to compare the usefulness of different sets of variable combinations among each other with respect to  group separation \citep{Akaboha2016, Bhutta2015, Kleinberg2019, Shinba2021, Nan2018}. In this study, we have only examined relative importance of variables to group separation through computing the descriptive discriminant coefficients.  Predictive discriminant analysis may provide information on how useful these variables are in distinguishing the two groups. Since predictive discriminant analysis in its initial version assumes multivariate normality of the data, several robust extensions have been developed for multivariate repeated measures data \citep{Brobbey2021, Brobbey2022}.
	\\
	With respect to the data and research question, quality of life measures (SF-36 PCS, SF-36 MCS, QOLIBRI) seem to include partially redundant information for distinguishing patients with uncomplicated and complicated mild TBI when the functional outcome (GOSE) and post-traumatic stress (PCL-5), depression (PHQ-9), and anxiety (GAD-7) are measured repeatedly. This outcome set has already been examined for cross-sectional sensitivity in preselected patient groups at 3, 6, and 12 months after TBI, respectively \citep{Steinbuechel2023}, suggesting that a reduced set of outcome measures would sufficiently discriminate between them, given the intercorrelation between the outcome measures from a clinical content perspective. Only then can the research findings make a valuable contribution to the development of clinical implications and tailored therapies. 
	
% \[...\]

	\vfill

\clearpage
%	 \bmhead{Supplementary information}
\textbf{Supplementary information} 

	\begin{figure}[htb!]
		\centering	
		\begin{subfigure}[b]{.32\linewidth}
			\includegraphics[width=\linewidth]{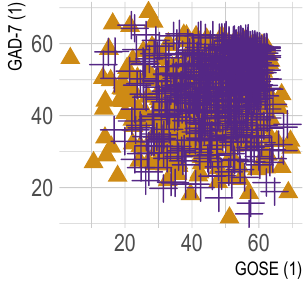}
		\end{subfigure}	
		\begin{subfigure}[b]{.32\linewidth}
			\includegraphics[width=\linewidth]{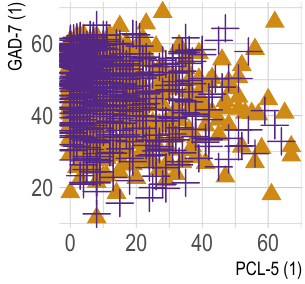}
		\end{subfigure}
		\begin{subfigure}[b]{.32\linewidth}
			\includegraphics[width=\linewidth]{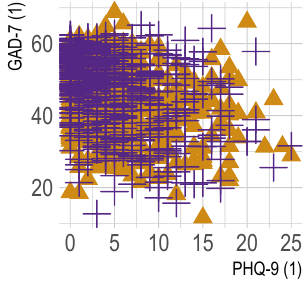}
		\end{subfigure}
		\begin{subfigure}[b]{.32\linewidth}
			\includegraphics[width=\linewidth]{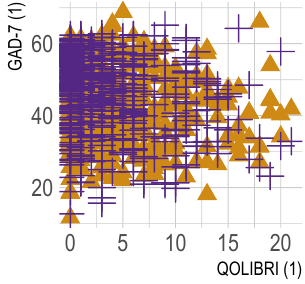}
		\end{subfigure}	
		\begin{subfigure}[b]{.32\linewidth}
			\includegraphics[width=\linewidth]{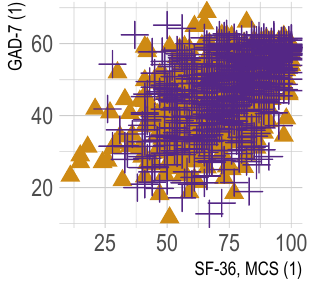}
		\end{subfigure}
		\begin{subfigure}[b]{.32\linewidth}
			\includegraphics[width=\linewidth]{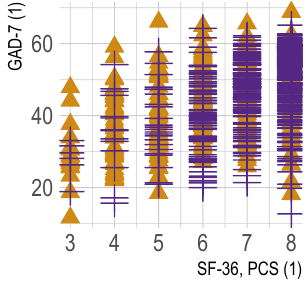}
		\end{subfigure}
		\begin{subfigure}[b]{.32\linewidth}
			\includegraphics[width=\linewidth]{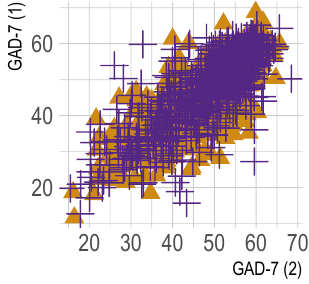}
		\end{subfigure}	
		\begin{subfigure}[b]{.32\linewidth}
			\includegraphics[width=\linewidth]{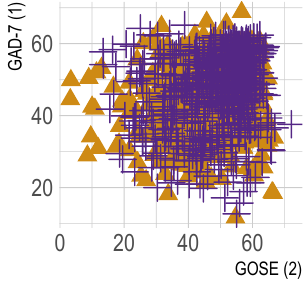}
		\end{subfigure}
		\begin{subfigure}[b]{.32\linewidth}
			\includegraphics[width=\linewidth]{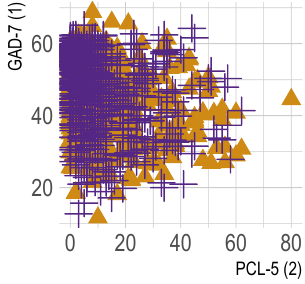}
		\end{subfigure}
		\begin{subfigure}[b]{.32\linewidth}
			\includegraphics[width=\linewidth]{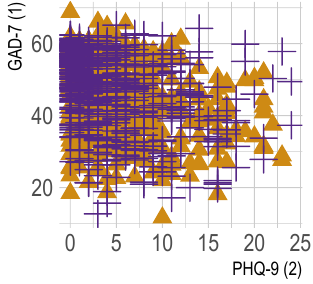}
		\end{subfigure}	
		\begin{subfigure}[b]{.32\linewidth}
			\includegraphics[width=\linewidth]{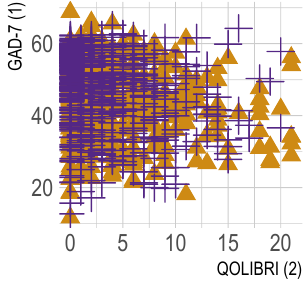}
		\end{subfigure}
		\begin{subfigure}[b]{.32\linewidth}
			\includegraphics[width=\linewidth]{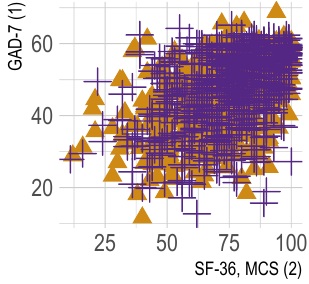}
		\end{subfigure}
		\begin{subfigure}[b]{.32\linewidth}
			\includegraphics[width=\linewidth]{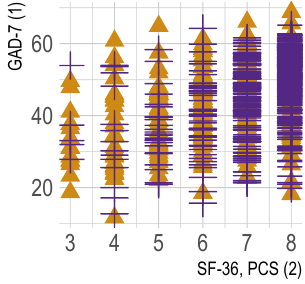}
		\end{subfigure}
		\begin{subfigure}[b]{.32\linewidth}
			\includegraphics[width=\linewidth]{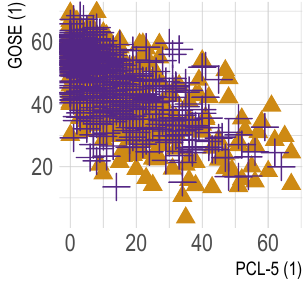}
		\end{subfigure}
		\begin{subfigure}[b]{.32\linewidth}
			\includegraphics[width=\linewidth]{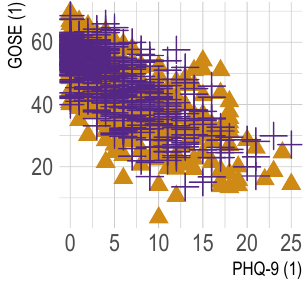}
		\end{subfigure}
		\begin{subfigure}[b]{.32\linewidth}
			\includegraphics[width=\linewidth]{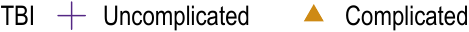}
		\end{subfigure}
		%	\caption{Picture of animals}
		%	\label{fig:animals}
	\end{figure}

%\clearpage

\begin{figure}
	\centering
	\begin{subfigure}[b]{.32\linewidth}
		\includegraphics[width=\linewidth]{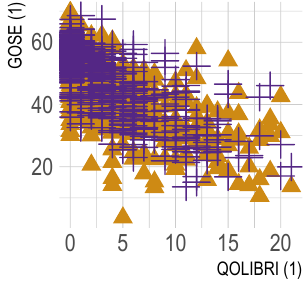}
	\end{subfigure}	
	\begin{subfigure}[b]{.32\linewidth}
		\includegraphics[width=\linewidth]{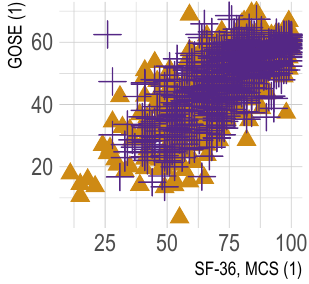}
	\end{subfigure}
	\begin{subfigure}[b]{.32\linewidth}
		\includegraphics[width=\linewidth]{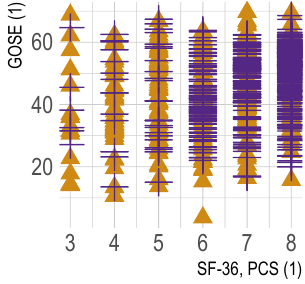}
	\end{subfigure}
	\begin{subfigure}[b]{.32\linewidth}
		\includegraphics[width=\linewidth]{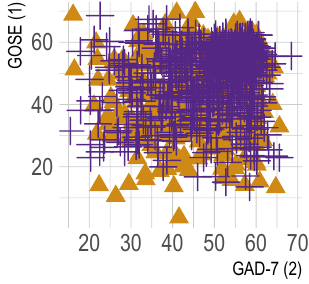}
	\end{subfigure}	
	\begin{subfigure}[b]{.32\linewidth}
		\includegraphics[width=\linewidth]{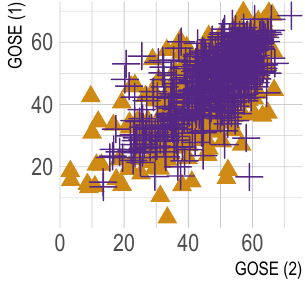}
	\end{subfigure}
	\begin{subfigure}[b]{.32\linewidth}
		\includegraphics[width=\linewidth]{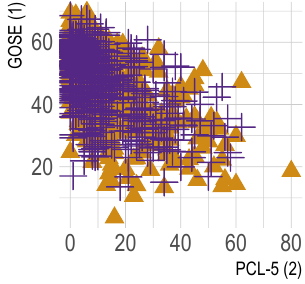}
	\end{subfigure}
	\begin{subfigure}[b]{.32\linewidth}
		\includegraphics[width=\linewidth]{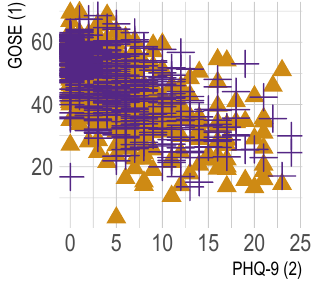}
	\end{subfigure}	
	\begin{subfigure}[b]{.32\linewidth}
		\includegraphics[width=\linewidth]{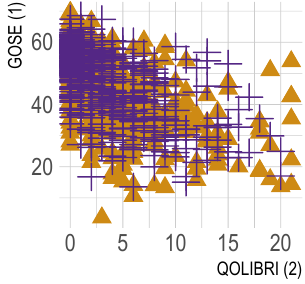}
	\end{subfigure}
	\begin{subfigure}[b]{.32\linewidth}
		\includegraphics[width=\linewidth]{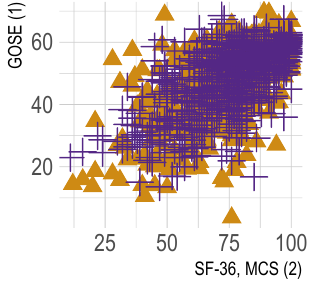}
	\end{subfigure}
	\begin{subfigure}[b]{.32\linewidth}
		\includegraphics[width=\linewidth]{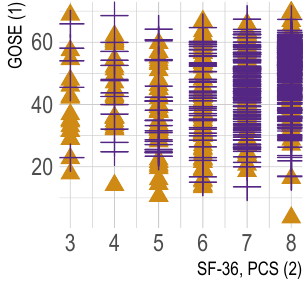}
	\end{subfigure}	
	\begin{subfigure}[b]{.32\linewidth}
		\includegraphics[width=\linewidth]{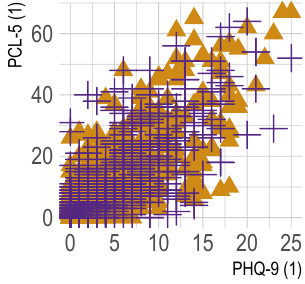}
	\end{subfigure}
	\begin{subfigure}[b]{.32\linewidth}
		\includegraphics[width=\linewidth]{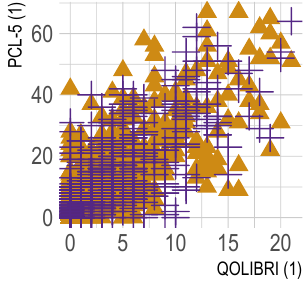}
	\end{subfigure}
	\begin{subfigure}[b]{.32\linewidth}
		\includegraphics[width=\linewidth]{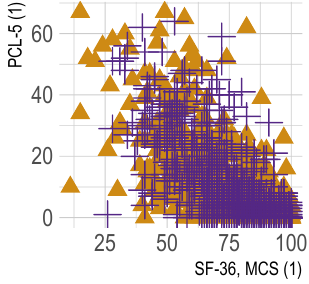}
	\end{subfigure}
	\begin{subfigure}[b]{.32\linewidth}
		\includegraphics[width=\linewidth]{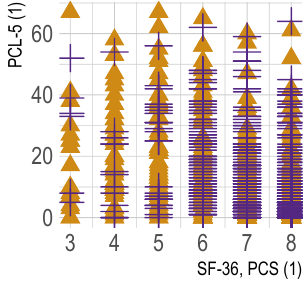}
	\end{subfigure}
	\begin{subfigure}[b]{.32\linewidth}
		\includegraphics[width=\linewidth]{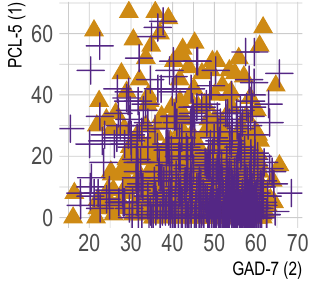}
	\end{subfigure}
	\begin{subfigure}[b]{.32\linewidth}
	\includegraphics[width=\linewidth]{figures/scatter/Fig_scatter_legend_bottom.pdf}
\end{subfigure}
	%	\caption{Picture of animals}
	%	\label{fig:animals}
\end{figure}

\begin{figure}
	\centering
	\begin{subfigure}[b]{.32\linewidth}
		\includegraphics[width=\linewidth]{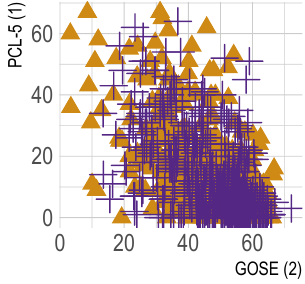}
	\end{subfigure}	
	\begin{subfigure}[b]{.32\linewidth}
		\includegraphics[width=\linewidth]{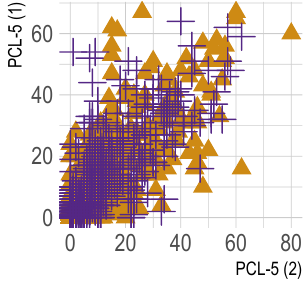}
	\end{subfigure}
	\begin{subfigure}[b]{.32\linewidth}
		\includegraphics[width=\linewidth]{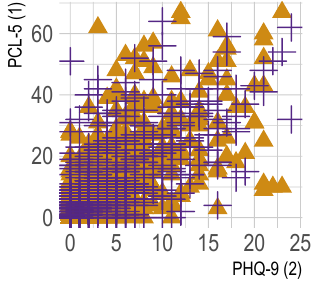}
	\end{subfigure}
	\begin{subfigure}[b]{.32\linewidth}
		\includegraphics[width=\linewidth]{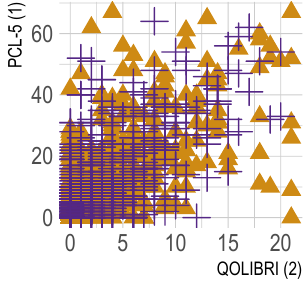}
	\end{subfigure}	
	\begin{subfigure}[b]{.32\linewidth}
		\includegraphics[width=\linewidth]{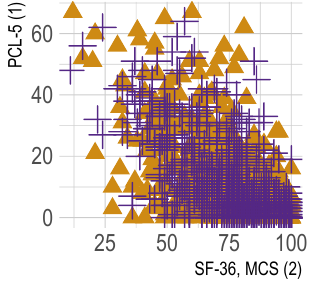}
	\end{subfigure}
	\begin{subfigure}[b]{.32\linewidth}
		\includegraphics[width=\linewidth]{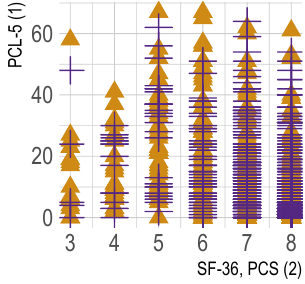}
	\end{subfigure}
	\begin{subfigure}[b]{.32\linewidth}
		\includegraphics[width=\linewidth]{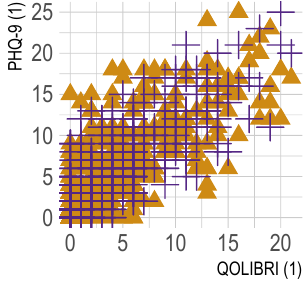}
	\end{subfigure}	
	\begin{subfigure}[b]{.32\linewidth}
		\includegraphics[width=\linewidth]{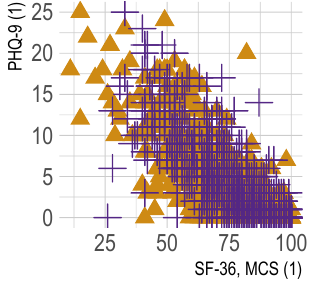}
	\end{subfigure}
	\begin{subfigure}[b]{.32\linewidth}
		\includegraphics[width=\linewidth]{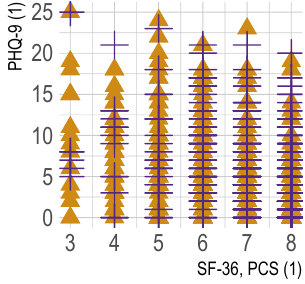}
	\end{subfigure}
	\begin{subfigure}[b]{.32\linewidth}
		\includegraphics[width=\linewidth]{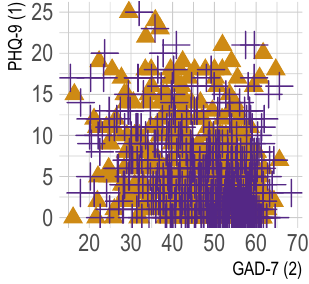}
	\end{subfigure}	
	\begin{subfigure}[b]{.32\linewidth}
		\includegraphics[width=\linewidth]{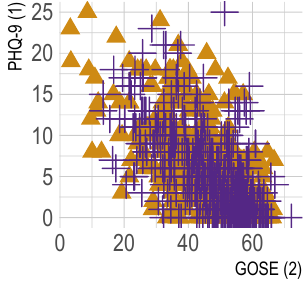}
	\end{subfigure}
	\begin{subfigure}[b]{.32\linewidth}
		\includegraphics[width=\linewidth]{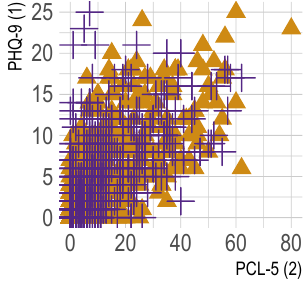}
	\end{subfigure}
	\begin{subfigure}[b]{.32\linewidth}
		\includegraphics[width=\linewidth]{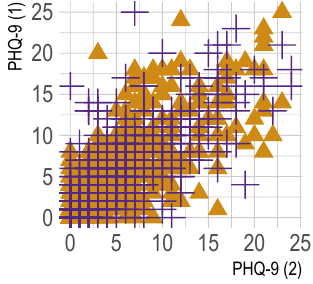}
	\end{subfigure}
	\begin{subfigure}[b]{.32\linewidth}
		\includegraphics[width=\linewidth]{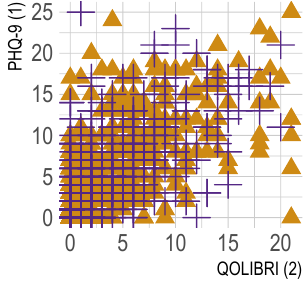}
	\end{subfigure}
	\begin{subfigure}[b]{.32\linewidth}
		\includegraphics[width=\linewidth]{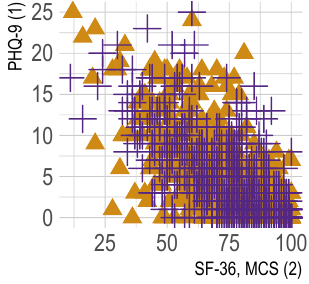}
	\end{subfigure}
	\begin{subfigure}[b]{.32\linewidth}
	\includegraphics[width=\linewidth]{figures/scatter/Fig_scatter_legend_bottom.pdf}
\end{subfigure}
	%	\caption{Picture of animals}
	%	\label{fig:animals}
\end{figure}

\begin{figure}
	\centering
	\begin{subfigure}[b]{.32\linewidth}
		\includegraphics[width=\linewidth]{figures/scatter/Fig_scatter_x15_x14.pdf}
	\end{subfigure}	
	\begin{subfigure}[b]{.32\linewidth}
		\includegraphics[width=\linewidth]{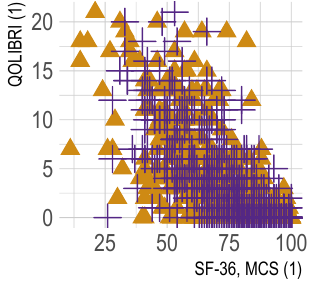}
	\end{subfigure}
	\begin{subfigure}[b]{.32\linewidth}
		\includegraphics[width=\linewidth]{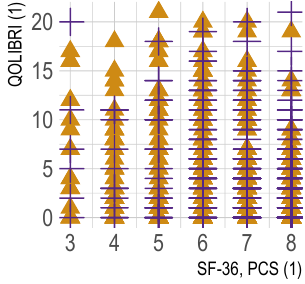}
	\end{subfigure}
	\begin{subfigure}[b]{.32\linewidth}
		\includegraphics[width=\linewidth]{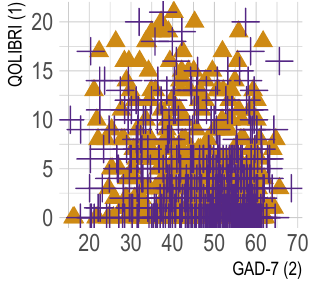}
	\end{subfigure}	
	\begin{subfigure}[b]{.32\linewidth}
		\includegraphics[width=\linewidth]{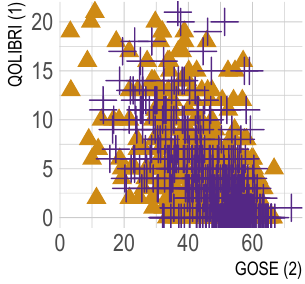}
	\end{subfigure}
	\begin{subfigure}[b]{.32\linewidth}
		\includegraphics[width=\linewidth]{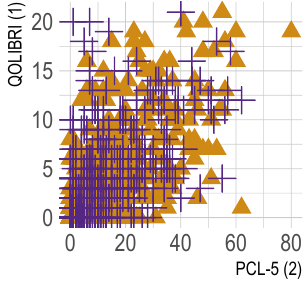}
	\end{subfigure}
	\begin{subfigure}[b]{.32\linewidth}
		\includegraphics[width=\linewidth]{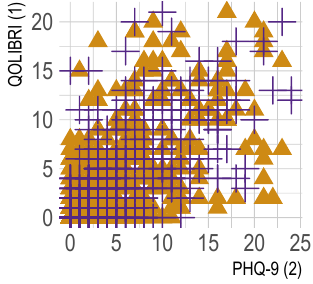}
	\end{subfigure}	
	\begin{subfigure}[b]{.32\linewidth}
		\includegraphics[width=\linewidth]{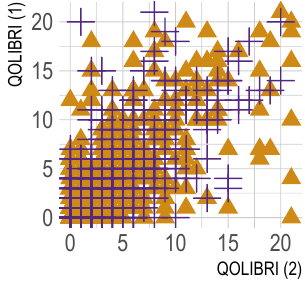}
	\end{subfigure}
	\begin{subfigure}[b]{.32\linewidth}
		\includegraphics[width=\linewidth]{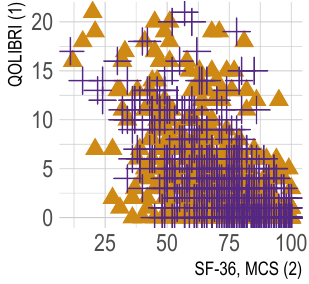}
	\end{subfigure}
	\begin{subfigure}[b]{.32\linewidth}
		\includegraphics[width=\linewidth]{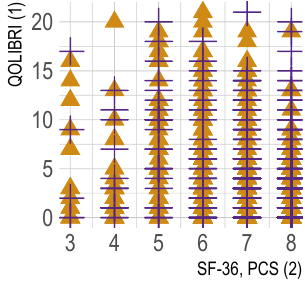}
	\end{subfigure}	
	\begin{subfigure}[b]{.32\linewidth}
		\includegraphics[width=\linewidth]{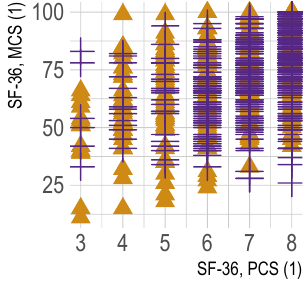}
	\end{subfigure}
	\begin{subfigure}[b]{.32\linewidth}
		\includegraphics[width=\linewidth]{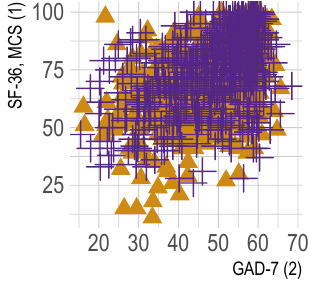}
	\end{subfigure}
	\begin{subfigure}[b]{.32\linewidth}
		\includegraphics[width=\linewidth]{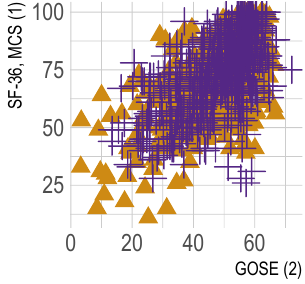}
	\end{subfigure}
	\begin{subfigure}[b]{.32\linewidth}
		\includegraphics[width=\linewidth]{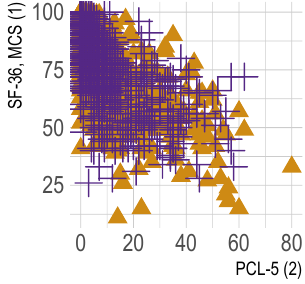}
	\end{subfigure}
	\begin{subfigure}[b]{.32\linewidth}
		\includegraphics[width=\linewidth]{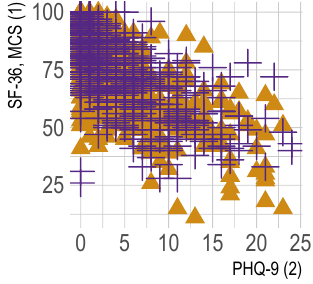}
	\end{subfigure}
	\begin{subfigure}[b]{.32\linewidth}
	\includegraphics[width=\linewidth]{figures/scatter/Fig_scatter_legend_bottom.pdf}
\end{subfigure}
	%	\caption{Picture of animals}
	%	\label{fig:animals}
\end{figure}

\begin{figure}
	\centering
	\begin{subfigure}[b]{.32\linewidth}
		\includegraphics[width=\linewidth]{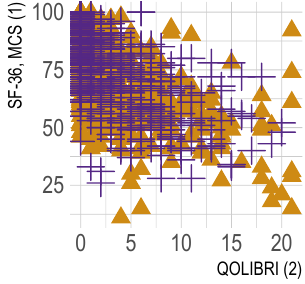}
	\end{subfigure}	
	\begin{subfigure}[b]{.32\linewidth}
		\includegraphics[width=\linewidth]{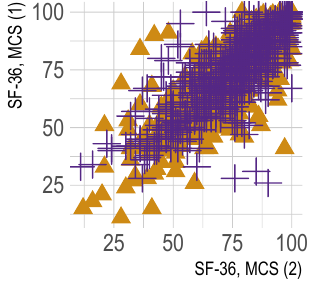}
	\end{subfigure}
	\begin{subfigure}[b]{.32\linewidth}
		\includegraphics[width=\linewidth]{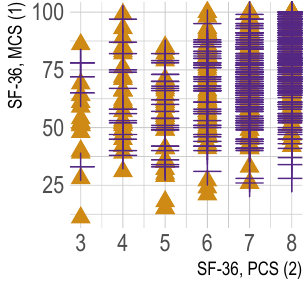}
	\end{subfigure}
	\begin{subfigure}[b]{.32\linewidth}
		\includegraphics[width=\linewidth]{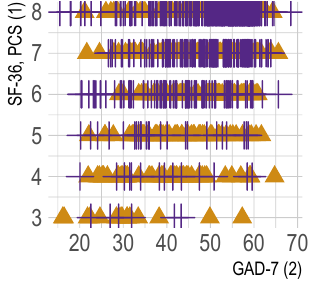}
	\end{subfigure}	
	\begin{subfigure}[b]{.32\linewidth}
		\includegraphics[width=\linewidth]{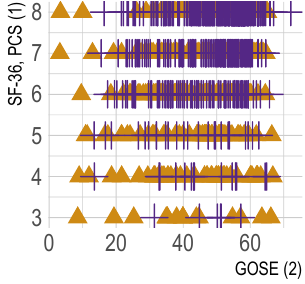}
	\end{subfigure}
	\begin{subfigure}[b]{.32\linewidth}
		\includegraphics[width=\linewidth]{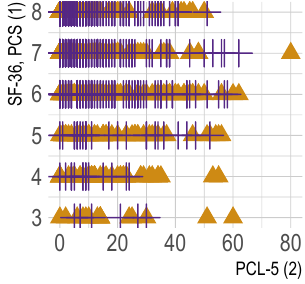}
	\end{subfigure}
	\begin{subfigure}[b]{.32\linewidth}
		\includegraphics[width=\linewidth]{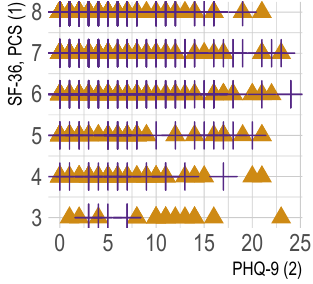}
	\end{subfigure}	
	\begin{subfigure}[b]{.32\linewidth}
		\includegraphics[width=\linewidth]{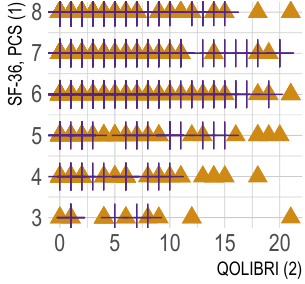}
	\end{subfigure}
	\begin{subfigure}[b]{.32\linewidth}
		\includegraphics[width=\linewidth]{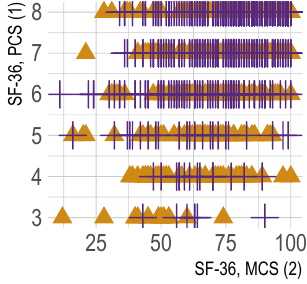}
	\end{subfigure}
	\begin{subfigure}[b]{.32\linewidth}
		\includegraphics[width=\linewidth]{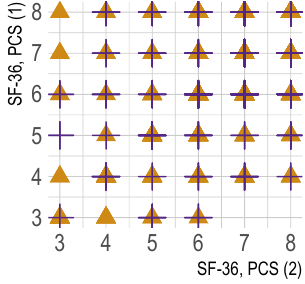}
	\end{subfigure}	
	\begin{subfigure}[b]{.32\linewidth}
		\includegraphics[width=\linewidth]{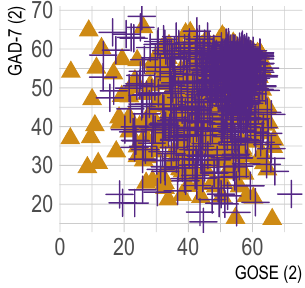}
	\end{subfigure}
	\begin{subfigure}[b]{.32\linewidth}
		\includegraphics[width=\linewidth]{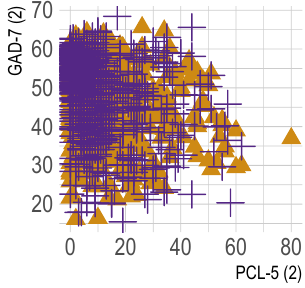}
	\end{subfigure}
	\begin{subfigure}[b]{.32\linewidth}
		\includegraphics[width=\linewidth]{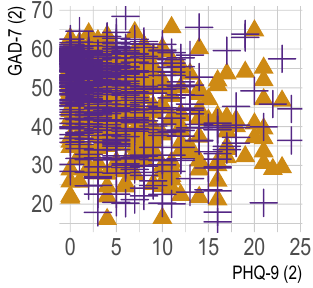}
	\end{subfigure}
	\begin{subfigure}[b]{.32\linewidth}
		\includegraphics[width=\linewidth]{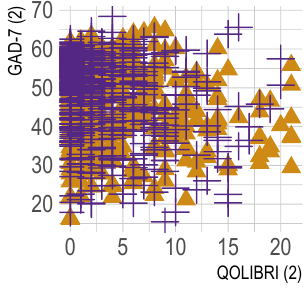}
	\end{subfigure}
	\begin{subfigure}[b]{.32\linewidth}
		\includegraphics[width=\linewidth]{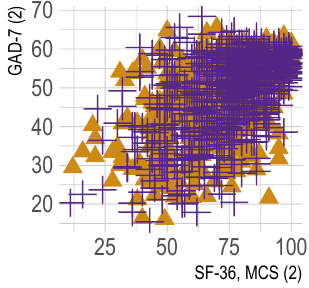}
	\end{subfigure}
	\begin{subfigure}[b]{.32\linewidth}
	\includegraphics[width=\linewidth]{figures/scatter/Fig_scatter_legend_bottom.pdf}
\end{subfigure}
	%	\caption{Picture of animals}
	%	\label{fig:animals}
\end{figure}

\begin{figure}
	\centering
	\begin{subfigure}[b]{.32\linewidth}
		\includegraphics[width=\linewidth]{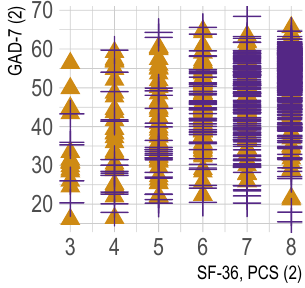}
	\end{subfigure}	
	\begin{subfigure}[b]{.32\linewidth}
		\includegraphics[width=\linewidth]{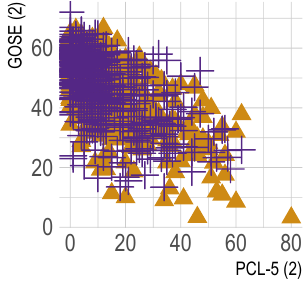}
	\end{subfigure}
	\begin{subfigure}[b]{.32\linewidth}
		\includegraphics[width=\linewidth]{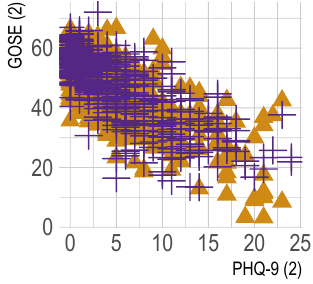}
	\end{subfigure}
	\begin{subfigure}[b]{.32\linewidth}
		\includegraphics[width=\linewidth]{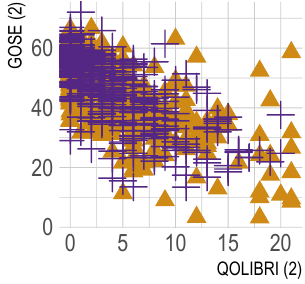}
	\end{subfigure}	
	\begin{subfigure}[b]{.32\linewidth}
		\includegraphics[width=\linewidth]{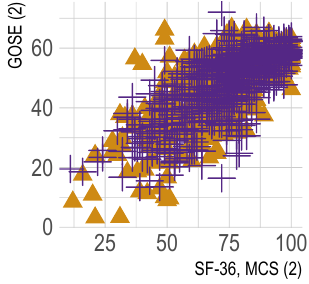}
	\end{subfigure}
	\begin{subfigure}[b]{.32\linewidth}
		\includegraphics[width=\linewidth]{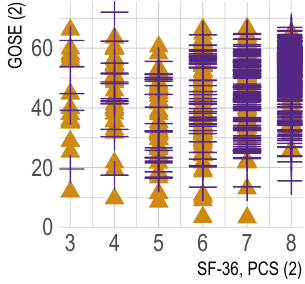}
	\end{subfigure}
	\begin{subfigure}[b]{.32\linewidth}
		\includegraphics[width=\linewidth]{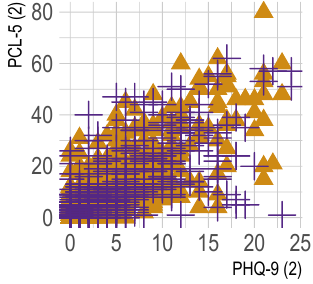}
	\end{subfigure}	
	\begin{subfigure}[b]{.32\linewidth}
		\includegraphics[width=\linewidth]{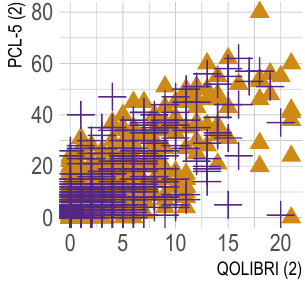}
	\end{subfigure}
	\begin{subfigure}[b]{.32\linewidth}
		\includegraphics[width=\linewidth]{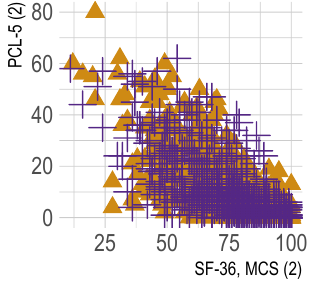}
	\end{subfigure}
	\begin{subfigure}[b]{.32\linewidth}
		\includegraphics[width=\linewidth]{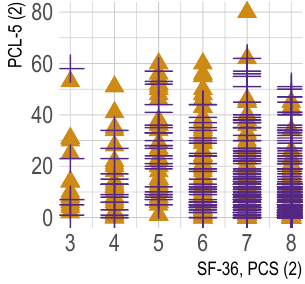}
	\end{subfigure}	
	\begin{subfigure}[b]{.32\linewidth}
		\includegraphics[width=\linewidth]{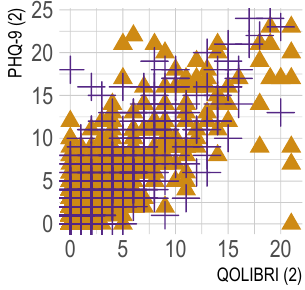}
	\end{subfigure}
	\begin{subfigure}[b]{.32\linewidth}
		\includegraphics[width=\linewidth]{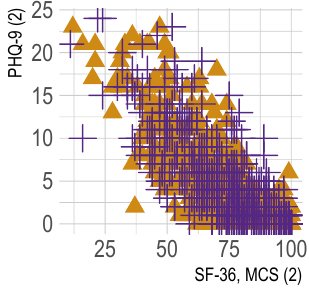}
	\end{subfigure}
	\begin{subfigure}[b]{.32\linewidth}
		\includegraphics[width=\linewidth]{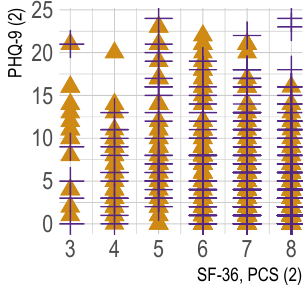}
	\end{subfigure}
	\begin{subfigure}[b]{.32\linewidth}
		\includegraphics[width=\linewidth]{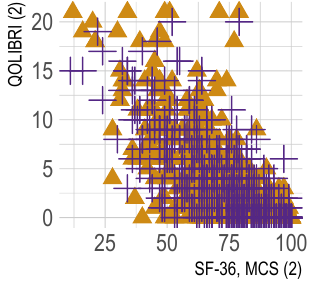}
	\end{subfigure}
	\begin{subfigure}[b]{.32\linewidth}
		\includegraphics[width=\linewidth]{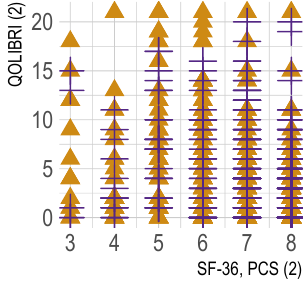}
	\end{subfigure}
	\begin{subfigure}[b]{.32\linewidth}
	\includegraphics[width=\linewidth]{figures/scatter/Fig_scatter_legend_bottom.pdf}
\end{subfigure}
	%	\caption{Picture of animals}
	%	\label{fig:animals}
\end{figure}

\thispagestyle{empty}
\pagebreak 

\begin{table}[htb!]
	\renewcommand\thetable{S 1}
	\caption{ \footnotesize Assessment of multicollinearity using scaled condition indices and scaled variance decomposition proportions after removal of multicollinear variables (SF-36 PCS, GAD-7, QOLIBRI).  Only (scaled) variance decomposition proportions $>$ .3 are shown.}
	\label{multicoll4}
	{\footnotesize
		\begin{tabular}{L{2.3cm}cccccccccc} 
			\toprule
			\multirow{2}{*}[-1.5em]{\parbox{2.3cm}{\small{(Scaled)\\ condition index}}}	& \multicolumn{8}{c}{\small{(Scaled) variance decomposition proportions}}& & \\[1ex] \cmidrule(rl){2-11}
			& \rot{SF-36, MCS (1)} &  \rot{PCL-5 (1)} & \rot{PHQ-9 (1)}    & \rot{GOSE (1)} & \rot{SF-36, MCS (2)} &  \rot{PCL-5 (2)} & \rot{PHQ-9 (2)}  & \rot{GOSE (2)} & &\\ \arrayrulecolor{gray!50}\cmidrule(lr){1-11} 
			2.31	 & . & . & . & . & . & . & . & .  &  & 			 \\ \arrayrulecolor{gray!50}\cmidrule(lr){2-11}
			5.94	 & . & . & . & . & . & . & . & .  &  &   			 \\ \arrayrulecolor{gray!50}\cmidrule(lr){2-11}	
			6.24	 & . & . & . & . & . & . & . & .  &  &   			 \\ \arrayrulecolor{gray!50}\cmidrule(lr){2-11}
			9.92	 & . & .58 & .55 & . & . & .5 & .48 & .   &  &  			 \\ \arrayrulecolor{gray!50}\cmidrule(lr){2-11}
			15.06	 & . & .  & . & . & . & .  & . & . &  &	 \\ \arrayrulecolor{gray!50}\cmidrule(lr){2-11}
			22.67	 & . & . & . & .47 & . & . & . &  .  &  & 	 \\ \arrayrulecolor{gray!50}\cmidrule(lr){2-11}
			25.82	 & . & . & . & . & . & . & . & .69  &  &  		 \\ \arrayrulecolor{gray!50}\cmidrule(lr){2-11}
			29.25	 & .66 & . & . & .31 & .7 & . & . & . &  &  		 \\ 				
			\arrayrulecolor{black}\bottomrule     
		\end{tabular}
	}
\end{table}
	
\begin{table}[htb!]
	\renewcommand\thetable{S 2}
	\caption{Standardized discriminant function coefficients (DFC) ordered by highest absolute value after removal of multicollinear variables SF-36 MCS, GAD-7, QOLIBRI.}
	\label{DFC_multcoll_suppl}
	
	\begin{tabular}{l} 
		\begin{tabular}{ll}  \toprule
			Variable & Stand. DFC \\ \cmidrule(rl){1-2}
			GOSE (1)     & -0.66    \\ % x17   
			GOSE (2)     & -0.62   \\ % x27         
			SF-36, PCS (2)     & -0.4  \\ %  x21    
			PCL-5 (2)     & -0.34   \\ % x23       
			PHQ-9 (2)     & -0.25   \\ % x24   
			PCL-5 (1)     & -0.17 \\  % x13      
			PHQ-9 (1)      & -0.12   \\  % x14         
			SF-36, PCS (1)     & -0.07   \\  % x11
			\arrayrulecolor{black}\bottomrule
			& \\
			& \\			
		\end{tabular}
	\end{tabular}
\end{table}

\begin{table}[htb!]
	\renewcommand\thetable{S 3}
	\caption{ \footnotesize Repeated measures (M)ANOVA results for the CENTER-TBI data after exclusion of different sets of multicollinear variables.
		MATS = modified ANOVA-type statistic, between-subject factor = TBI severity (uncomplicated and complicated mTBI), within-subject factor = time (time points 3 and 6 months after TBI),
		$p$ = $p$-value based on parametric bootstrapping, bold $p$-values are significant at $\alpha$ = 0.05 for MANOVA-RM. Abbreviations: mTBI = mild traumatic brain injury;  SF-PCS = Short Form (36) Health Survey (physical component score); SF-MCS = Short Form (36) Health Survey (mental component score); GAD-7 = Generalized Anxiety Disorder questionnaire; QOLIBRI = Quality of Life after Brain Injury; .}
	\label{RMMANOVA_coll}
	{\footnotesize
		\begin{tabular}{cllllll} 
			&&& \\
			\textbf{\parbox{1.5cm}{Analysis\\}}  &   \textbf{\parbox{4.5cm}{Excluded multicollinear\\dependent variables}} 		&  \textbf{\parbox{3cm}{Independent variable \\}}  &  \textbf{\parbox{2.5cm}{MATS \\}}  & \textbf{df1} & \textbf{df2} & \textbf{$p$-value}    \\ \cmidrule(lr){1-7}
			&	\parbox{3.5cm}{(a) SF-36 PCS, QOLIBRI}	&	mTBI       	&  165.021 & \textminus  & \textminus &  \textbf{\textless{}.001} \\
			&								&	Time points      		&  19.68  & \textminus  & \textminus  &\textbf{\textless{}.001}  \\ 
			&								&	mTBI:Time points 		&  1.377 & \textminus  & \textminus  & .398    \\ \cmidrule(lr){2-7} 
			MANOVA		  &	\parbox{3.5cm}{(b) SF-36 PCS, SF-36 MCS}	&	mTBI       	&  173.359 & \textminus  & \textminus & \textbf{\textless{}.001} \\
			RM 			  				&							&	Time points      		&   18.64  & \textminus  & \textminus  & \textbf{\textless{}.001}  \\ 
			&							&	mTBI:Time points 		&  1.172 & \textminus  & \textminus  & .46    \\ \cmidrule(lr){2-7} 
			&	\parbox{4.5cm}{(c) SF-36 MCS, GAD-7, QOLIBRI}	&	mTBI       	&  156.369 & \textminus  & \textminus  & \textbf{\textless{}.001} \\
			&								&	Time points      		&  29.261 & \textminus  & \textminus & \textbf{\textless{}.001}  \\ 
			&								&	mTBI:Time points 		&  1.505 & \textminus  & \textminus & .227    \\ \arrayrulecolor{black}\cmidrule(lr){1-7} 
			
		\end{tabular}
	}
\end{table}

	\bigskip

	\clearpage
	\bibliographystyle{apalike}

	\bibliography{sn-bibliography}% common bib file

%\printbibliography

%	\input sn-article.bbl

\end{document}